\documentclass[12pt]{article}
\usepackage[top=26mm, bottom=40mm, left=26mm, right=26mm]{geometry}
\usepackage{graphicx} 
\usepackage{amsfonts}
\usepackage{amsmath}
\usepackage{xcolor}
\usepackage[hypcap=false]{caption}
\usepackage{multirow}
\usepackage{tablefootnote}
\usepackage{float}
\usepackage{bbold}
\usepackage{hyperref}
\usepackage{physics}
\usepackage{ragged2e}
\usepackage[citestyle=numeric,bibstyle=numeric,
giveninits=true,doi=false,url=false,isbn=false,
parentracker=false,
backend=biber,
maxnames=10,
date=year,
sorting=none]{biblatex} 
\makeatletter

\renewbibmacro{in:}{}
\DeclareFieldFormat{doilink}{%
  \iffieldundef{doi}%
  {#1}%
  {\href{https://doi.org/\thefield{doi}}{#1}}
}
\DeclareBibliographyDriver{article}{%
  \usebibmacro{bibindex}%
  \usebibmacro{begentry}%
  \usebibmacro{author/translator+others}%
  \setunit{\printdelim{nametitledelim}}\newblock
  \usebibmacro{title}%
  \newunit
  \printlist{language}%
  \newunit\newblock
  \usebibmacro{byauthor}%
  \newunit\newblock
  \usebibmacro{bytranslator+others}%
  \newunit\newblock
  \printfield{version}%
  \newunit\newblock
  \usebibmacro{in:}
  \iffieldundef{journaltitle}
   {\printtext[journaltitle]{Preprint} \mkbibparens{\printdate}}
   {\printtext[doilink]{%
    \usebibmacro{journal+issuetitle}%
    \newunit
    \usebibmacro{byeditor+others}%
    \newunit
    \usebibmacro{note+pages}%
    \newunit\newblock
    \iftoggle{bbx:isbn}
      {\printfield{issn}}
      {}%
   }
   }
  \newunit\newblock
  \usebibmacro{doi+eprint+url}
  \newunit\newblock
  \usebibmacro{addendum+pubstate}%
  \setunit{\bibpagerefpunct}\newblock
  \usebibmacro{pageref}%
  \newunit\newblock
  \iftoggle{bbx:related}
    {\usebibmacro{related:init}%
     \usebibmacro{related}}
    {}%
  \usebibmacro{finentry}}

\letbibmacro{nakedEprint}{eprint}
\renewbibmacro*{eprint}{%
  \iffieldundef{eprint}
    {}
    {\iffieldundef{journaltitle}
      {\usebibmacro{nakedEprint}}
      {\printtext{{\usebibmacro{nakedEprint}}}}} 
}
\DeclareFieldFormat{eprint:arxiv}{%
  \ifhyperref
    {\href{https://arxiv.org/\abx@arxivpath/#1}{%
       \nolinkurl{arXiv:#1}%
       \iffieldundef{eprintclass}
         {}
         {\addspace{\mkbibbrackets{\thefield{eprintclass}}}}}}  
    {\nolinkurl{arXiv:#1}%
     \iffieldundef{eprintclass}
       {}
       {\addspace{\mkbibbrackets{\thefield{eprintclass}}}}}}   

\makeatother

\def\L{\Lambda}
\def\Z{\mathbb{Z}}
\def\CP{\mathbb{CP}}
\def\WCP{\mathbb{WCP}}
\def\N{\mathcal{N}}
\def\W{\mathcal{W}}
\def\C{\mathbb{C}}
\newcommand{\lz}{[\Lambda_0]}
\newcommand{\lo}{[\Lambda_1]}
\newcommand{\lt}{[\Lambda_2]}

\numberwithin{equation}{section}

\begin{document}

\begin{titlepage}

\vspace*{-2cm} 

\vspace*{0.8cm} 
\begin{center}
{\Huge Orientifolds of Gepner models \\
\vspace{0.3cm}
without K{\"a}hler moduli}\\

\vspace*{0.5cm}
Miguel Morros$^{1}$, Johannes Walcher$^{2}$ and Qi You$^{3}$\\

\vspace*{1.0cm} 

$^1$ {\it Institut de Physique Th\'eorique, Universit\'e Paris Saclay, CEA, CNRS Orme des Merisiers, 91191 Gif-sur-Yvette CEDEX, France}\\[4mm]

$^2$ {\it Institute for Mathematics and Institute for Theoretical Physics Ruprecht-Karls-Universit{\"a}t Heidelberg, 69120 Heidelberg, Germany}\\[4mm]

$^3$ {\it Mitchell Institute for Fundamental Physics and Astronomy, Texas A$\&$M University College Station, TX 77843, USA}\\[3mm]

{\tt {miguel.morros@ipht.fr, walcher@uni-heidelberg.de, qi\_you@physics.tamu.edu}} \\

\vspace{2cm}
\small{\bf Abstract} \\[3mm]\end{center}

One of the main challenges in string theory Calabi-Yau compactifications to four dimensions is the stabilization of the massless complex structure and K\"ahler moduli. In type IIB string theory, complex structure moduli can be stabilized perturbatively by turning on fluxes on the internal space, while there is no perturbative mechanism for K\"ahler moduli stabilization. Since every Calabi-Yau manifold has at least one K\"ahler modulus (the overall volume), there is no hope to stabilize all moduli perturbatively. A way out is given by Landau-Ginzburg/Gepner models string vacua which can have no K\"ahler moduli. To identify the most promising candidates for fully stabilized perturbative string vacua, we provide an exhaustive list of Landau-Ginzburg orientifold models with no K\"ahler moduli, and compute for each model the number of complex structure moduli together with the tadpole charge. From this, we can identify which of these models are genuine candidates for phenomenologically relevant string vacua.

\end{titlepage}

\tableofcontents

\newpage

\section{Introduction}
\label{sec:intro}

The study of string compactifications on Gepner models \cite{Gepner:1987qi,Gepner:1987vz} is fundamentally enriched by the Landau-Ginzburg (LG)/Calabi-Yau (CY) correspondence \cite{Greene:1988ut}, which provides a powerful tool for analyzing the physics of non-geometric vacua. Since geometric compactifications often lead to plenty of massless scalars in a low-dimensional effective theory, which is far from being phenomenologically interesting, non-geometric models with moduli stabilization have attracted increasing attention. The term ``non-geometry'' is broadly used with various meanings in the literature. For example, it may refer to models that cannot be described in terms of any (pseudo-)Riemannian geometry, models that are constructed by asymmetric simple current extensions of the Gepner CFT \cite{Blumenhagen:2016axv,Blumenhagen:2016rof} or those that can only be defined via T-duality but still may not be T-dual to any (pseudo-)Riemannian geometry \cite{Plauschinn:2018wbo}. In the context of this paper, being non-geometric means that the Calabi-Yau associated with a Gepner model does not have K\"ahler moduli. That being said, it is mirror dual to some rigid Calabi-Yau manifold. Flux compactification then serves as a powerful mechanism for moduli stabilization, as background fluxes generate a scalar potential \cite{Gukov:1999ya} that can fix at a perturbative level otherwise massless complex structure moduli as well as the dilaton. Besides, the GVW superpotential itself is non-renormalized even in this kind of non-geometric case, meaning that it does not receive either perturbative or non-perturbative corrections \cite{Becker_2007}. Absence of K\"ahler moduli means we do not have to invoke non-perturbative effects, such as gaugino condensation or D-brane instantons as in KKLT or large volume scenario \cite{Kachru_2002, Balasubramanian:2005zx}, to stabilize them. For the theory to be globally consistent, the (positive) D-brane charges generated by the fluxes (and background D-branes, if any) thus requires an orientifold projection to cancel the RR anomaly, known as the tadpole cancellation condition, which has only been well understood in the case of geometric flux compactifications. However, since D-branes and orientifolds in LG/Gepner models have been extensively studied \cite{Kapustin_2003, Kapustin_2004, Brunner_2004, Brunner_2007, Hori_2008} in the past two decades, we shall apply some techniques from them to our non-geometric case.

To systematically map out the landscape of these non-geometric vacua, it is imperative to perform an exhaustive classification of B-type orientifolds and compute the corresponding O-plane charges across the entire family of $h^{1,1}=0$ Gepner models. In this paper, we will follow the ideas in \cite{Becker_2007}, present all the possible orientifolds for all of the Gepner models with $h^{1,1}=0$, and compute the charges of their associated O-planes. Additionally, we will compute the tadpole charge of the O-planes of all models with only A-type modular invariants. The goal is to give tadpole cancellation conditions for flux compactification on B-type orientifolds of Landau-Ginzburg/Gepner models string vacua. Achieving this global classification, however, requires confronting a long-standing subtlety: the non-uniqueness of the LG/CY correspondence.

In \cite{Witten:1993yc}, Witten proposed that the LG theory and the sigma models on the associated CY geometry can be understood as two phases in some Gauged Linear Sigma Model (GLSM). From his viewpoint, it is believed that there exists a particular family of superconformal field theories interpolating from the IR CFT of the Landau-Ginzburg theory to that of the Calabi-Yau compactification. This gives a way to interpret the fluxes in a non-geometric background and requires some identification between Dolbeault cohomology classes of the Calabi-Yau and chiral primaries (or, equivalently under spectral flow, Ramond ground states) in the Gepner CFT. 
Some more general cases of LG/CY correspondence were also explored in \cite{Witten:1993yc, Gu:2023hgm, Sethi:1994ch}. However, it is worth noting that the LG/CY correspondence is not unique in some sense. In \cite{FUCHS1989317} and \cite{Fuchs:1989yv}, the authors pointed out ambiguities in identifying Calabi-Yau geometries and Gepner models with various affine invariants, arguing that there may be several possibilities to construct a Calabi-Yau when the system can be factorized into $c=3$ and $c=6$ pieces. Their argument was purely based on the heuristic method proposed in \cite{Greene:1988ut}, regarding manipulations of the LG superpotentials in the path integral to give rise to delta-functions that define a hyperplane in some (weighted) projected space. We will see models with such ambiguities later in this paper. On the one hand, such correspondence may depend on the particular GLSM one considers, and even in the same model, the path from the small volume to large volume regimes on the moduli space matters. Things become even more interesting when fluxes, orbifolds and orientifolds come into play. For example, it was shown in \cite{Becker_2007} that the contribution by fluxes to the tadpole is not invariant under small volume monodromies and an orientifold projection may eliminate some of the possible monodromies. Some of such examples will be discussed in \autoref{sec:orientifolds_gepmod} of this paper.

There are in total 12 non-geometric Gepner models with $h^{1,1} = 0$ \cite{Fuchs:1989yv}, $8$ of them involving D-type modular invariants. Two of the ``A-type invariants only'' models, namely the $1^9$ and $2^6$ models\footnote{This notation will be explained later.}, have been revisited a lot recently \cite{becker2023fluxes19landauginzburgmodel, Becker:2024ayh, Becker_2024_higherorder, chen2025symmetriesmtheorylikevacuadimensions}. Tadpole cancellation conditions were derived in \cite{Becker_2007} and many supersymmetric Minkowski, AdS and non-supersymmetric dS vacua with tachyons have been found in the literature aforementioned. Candidates of 4D Minkowski vacua with all moduli stabilized have been discovered recently in \cite{Becker:2024ayh}, obtained from a worldsheet treatment of type IIB flux compactification on a B-type orientifold of the $2^6$ Gepner model. Solutions with massless moduli that get lifted to full stabilization upon including higher order corrections are also found immediately\footnote{In our paper, we have found a different tadpole bound than \cite{Becker_2007} by carefully studying the geometry when CFT ``trivial" factors are added. It turns out that the flux configurations that stabilize all moduli found in \cite{Becker:2024ayh, Becker_2024_higherorder} have a flux charge that is larger than the O-plane charge we find, and some conclusions there request further justifications.} \cite{Becker_2024_higherorder}. The linear behavior of the tadpole conjecture was explicitly shown, although a violation of the value of the coefficient proposed in the first formulation of the tadpole conjecture \cite{Bena:2020xrh} was observed\footnote{Another example of violation in F-theory compactification has been found in \cite{Lust:2022mhk}, but with supersymmetry breaking solutions.}. Evidence showing that the massless Minkowski conjecture does not hold beyond supergravity was also presented. All this work suggests that non-geometric models deserve further investigations, both from a formal and a phenomenological perspective. In finding string vacua of these non-geometric Gepner models, most of the aforementioned papers highly depend on the validity of the non-renormalization property of the LG superpotential, which is believed to be true even in the non-geometric case. Nonetheless, an alternative argument to support the existence of all such solutions purely based on the symmetry of the theory has been proposed in \cite{chen2025symmetriesmtheorylikevacuadimensions}. More discussions from this aspect will be given in \autoref{sec:fluxquant}. Aside from the two models mentioned above, others feature more subtleties and may require further techniques to be well treated. One key aspect is that there can also be complex structure moduli coming from twisted sectors, and to stabilize them is not as simple. In \cite{Ishiguro:2024coq}, the authors derived the effective type IIB action of twisted moduli by utilizing a mirror symmetry technique and then numerically verified that the supersymmetric AdS vacua they found with a stable twisted modulus are consistent with two swampland conjectures. Additionally, D-type Gepner models often come with a more complicated geometric setting and the matrix factorization description \cite{Hori_2004, Walcher_2005} of the permutation B-branes involve a highly non-trivial algebraic treatment.

The paper will be organized as follows. In \autoref{sec:review_Gepner}, we review the basics of ADE-type Gepner models as a coset theory and we count the number of complex structure moduli from both untwisted and twisted sectors (see also \autoref{app:RRgs}). In \autoref{sec:B-branes_MF}, we describe the LG B-branes in the standard matrix factorization language and show its applications to the A-type Gepner models\footnote{In fact, we need A-branes to support fluxes, but it is more convenient to compute things in the mirror model, where they become B-branes and can be perfectly described by the matrix factorization language, though a bit more categorical.}. In particular, we use $1^3\times4^3_A$ model as an example with plenty of details. Then, to have an understanding of the ``dimension'' of D-branes (in particular, D3-branes that can support fluxes), we need a description from a more geometric perspective. Since the branes can be described using categories by the dual exceptional collection of some line bundles, restricted to the ambient space \cite{Douglas_2001, ASPINWALL_2005}, we can compute their Chern characters to find out all the (internal) D0-branes that can support 3-form fluxes. In \autoref{sec:orientifolds_gepmod}, we give the definition of B-type orientifolds and that of the associated B-parities. We also present (see \autoref{app:Oplanes} as well)  all the possible O-planes in various non-geometric A/D-type\footnote{There are no E-type non-geometric Gepner models within the scope of our considerations \cite{Fuchs:1989yv}.} Gepner models and compute their homology classes expressed in terms of the D-brane classes (for K-theory class descriptions, see \cite{Witten:1998cd, Braun:2005eg}). In \autoref{sec:fluxquant}, we show the explicit tadpole cancellation conditions for the $1^7\times4_A$ and $1^3\times4^3_A$ models and give some comments about flux quantizations. In the second part of this section, we discuss moduli stabilization of the theory and the role that enhanced (discrete) symmetry can play in more details. Some open questions will be discussed in the end.

\section{Review of Gepner models}

\label{sec:review_Gepner}

\subsection{CFT and Landau-Ginzburg descriptions}

Gepner models provide a worldsheet description (as an SCFT) of string compactifications, in particular, the internal space is chosen to be described as an orbifold of tensor product of $\mathcal{N}=2$ minimal models with total central charge $9$ (for compactifications to four space-time dimensions). Spacetime supersymmetry further requires GSO projection of the theory, which may also be considered as an orbifold by $(-1)^F$, where $F$ is the mod $2$ Fermion number. In most cases, Gepner models are in the same moduli space as Calabi-Yau compactifications. In the language of 2D gauged linear sigma model with $U(1)$ gauge group, the infinitely negative limit of the real part of the FI-theta parameter corresponds to an Orbifold of an LG model. In the IR, this LG model flows to a Gepner model. Linear sigma models on a Calabi-Yau manifold correspond to the low energy limit of the other limit of the real part if the FI-theta parameter, namely being infinitely positive \cite{Brunner_2007}. Since FI-theta parameter is related to the complexified K\"ahler class in the large volume limit, the Gepner model can thus be analytically connected to the Calabi-Yau compactification in some large volume regime, depending on the connectivity of the K\"ahler moduli space. However, in some other cases, there may not exist a Calabi-Yau manifold (in the geometric sense) that is connected to a Gepner model, these models are sometimes called non-geometric models, and may only be studied from the worldsheet perspective.

The building blocks of Gepner models are the $\mathcal{N}=(2,2)$ superconformal level-$k$ ($k=0,1,2,...$) minimal models $M_k$\footnote{When $k=0$, this is a trivial minimal model with $c=0$, which only has a vacuum state in its Hilbert space.}. They can be realized as the following coset model,
\begin{equation}
    \frac{\hat{\mathfrak{su}}(2)_k\times\hat{\mathfrak{u}}(1)_2}{\hat{\mathfrak{u}}(1)_{k+2}}
\end{equation}
with central charge
\begin{equation}
    c = \frac{3k}{k+2}.
\end{equation}
When these minimal models are tensored together to build a Gepner model, the central charges must add up to $c=9$, leading to a finite number of 168 possible ways \cite{Fuchs:1989yv} (without taking into account various affine invariants, see below) to arrange minimal models to build a $D=4+6$ string vacuum. The Gepner model $M_{k_1}^{\otimes n_1} \otimes ... \otimes M_{k_r}^{\otimes n_r}$ is simply noted $k_1^{n_1} \times ... \times k_r^{n_r}$. Among those 168 Gepner models, 12 models (counting different affine invariants this time) lie in our case of interest since they lead to string compactifications with $h^{1,1}=0$. The list is the following:
\begin{itemize}
    \item $1^9$
    \item $1^7 \times 4_A$ ; $1^7 \times 4_D$
    \item $1^5 \times 4_D \times 4_A$ ; $1^5 \times 4_D^2$
    \item $1^3 \times 4_A^3$ ; $1^3 \times 4_D^2 \times 4_A$ ; $1^3 \times 4_D^3$
    \item $2^6$
    \item $1 \times 4_D \times 4_A^3$ ; $1 \times 4_D^3 \times 4_A$ ; $1 \times 4_D^4$
\end{itemize}

The highest weight representations $\Phi_{m,s}^l$ of the level-$k$ minimal models are labeled by the numbers $(l,m,s)$, satisfying $0\leq l\leq k$ and $l+m+s\in 2\mathbb{Z}$. $m$ and $s$ are respectively defined modulo $2k+4$ and $4$, and are often set to satisfy $-k-1\leq m\leq k+2,~s=-1,0,1,2$ by a field identification\footnote{Note that there exist states that cannot be brought into such standard range. In the $NS$-sector, the conformal weight and $U(1)$ charge of states labeled by $(l,-l,2)$ with $l>0$ should be computed by setting $s=2$ instead of $-2$, while the state $(0,0,2)$ has $h=\frac{3}{2}$ and $q=1$ \cite{Blumenhagen:2013fgp}.}
\begin{equation}
    \Phi_{m,s}^l\sim\Phi_{m+k+2,s+2}^{k-l}.
\end{equation}
Their conformal weights and U(1) R-symmetry charges are given by
\begin{equation}
    h_{m,s}^l=\frac{l(l+2)-m^2}{4(k+2)}+\frac{s^2}{8},~~~q_{m,s}^l=-\frac{m}{k+2}+\frac{s}{2}.
\end{equation}
A particular class of fields the NS sector of $\N=2$ SCFT are the (anti-)chiral primary fields. They are defined by
\begin{equation}
    G^{+(-)}_{-1/2} \Phi_{m,s}^l = 0,
\end{equation}
where $G^\pm$ are the two supercurrents of the $\N=2$ super-Virasoro algebra and they satisfy 
\begin{equation}
    h_{m,s}^l = \frac{|q_{m,s}^l|}{2}.
\end{equation}
They are labeled by $(l,-l,0)$ and are isomorphic under spectral flow to the (degenerate) Ramond ground states, labeled by $(l,-l-1,-1)$. The chiral primary fields form a Dolbeault cohomology with the first modes of the supercurrents as operators, and in the context of LG/CY compactification, they are identified with the basis of complex harmonic forms of the compact space. In the following we will focus on the computation of the number of such chiral primaries/RR ground states as they provide the most direct access to the Hodge numbers of the compact space.

The modular invariant partition function of $\hat{\mathfrak{su}}(2)_k$ (and thus of the coset model) can be classified in terms of ADE-type \cite{Fuchs:1989yv, CAPPELLI1987445, Qiu:1987ux}. Hence most Gepner models can have several variations depending on the modular invariant partition function chosen for its submodels. In models with $h^{1,1}=0$, it happens that only A-type models at level $k=1,2,4$ and D-type model at level $k=4$ shall appear\footnote{For $k=2$, the A- and D-type modular invariants happen to be the same.}. We give the corresponding $\hat{\mathfrak{su}}(2)_k$ partition functions in terms of the characters of the $\hat{\mathfrak{su}}(2)_k$ highest weight representations
\begin{align}
    Z_{\mathrm{A}_{k+1}} & = \sum_{l=0}^k |\chi_l|^2, \\
    Z_{\mathrm{D}_{2j+2}} & = \sum_{l=0}^{j-1} |\chi_{2l} + \chi_{k-2l}|^2 + 2 |\chi_{k/2}|^2 ~~~ (k=4j), \label{eq:ZD} \\
    Z_{\mathrm{D}_{2j+1}} & = \sum_{l=0}^{k/2} |\chi_{2l}|^2 + \sum_{l=0}^{k/2-1} \chi_{2l+1} \bar{\chi}_{k-2l-1} ~~~ (k=4j-2).
\end{align}

The superpotential of the LG model that flows to a given Gepner model in the IR \cite{Greene:1988ut,Vafa:1988uu} is also given by the level and the type of modular invariant of each submodel. The correspondences for each minimal model are listed in \autoref{ADE LG superpotentials}, and the LG superpotential is simply the sum of all submodels. In the framework of LG/CY correspondence, the CY manifold is defined as a (product of) hypersurface in (weighted) projective space, defined by the (set of) homogeneous equations $\mathcal{W} = 0$ \footnote{More precisely, the CY space is defined by a sole equation when the number of minimal factors is $r=5$ (or $r=4$ plus a trivial minimal model). If $r\geq6$, direct product of spaces for subtheories should be considered instead \cite{Fuchs:1989yv}.}, where $\mathcal{W}$ is the worldsheet superpotential. Also note that one or more trivial minimal models shall be added when the theory has an even number of A-type minimal models \cite{Lynker:1990gh}. We omit them in the expression of LG superpotentials and RR ground states for conciseness of notations but they should be restored when computing B-brane charges, intersection matrix, O-plane charges, etc. (see the examples in \autoref{sec:B-branes_MF} and \autoref{sec:orientifolds_gepmod}).

As mentioned above, we still need to take an orbifold of the LG/Gepner model to construct a string vacuum. The orbifold group is $\mathbb{Z}_H$ where $H$ is the degree of the LG superpotential. In the A-series of Gepner models $H=\mathrm{lcm}\{k_i+2\}$, while they may differ by a factor of two for D-type Gepner models, more precisely, $H=\mathrm{lcm}\{k_i+2\}/2$ when there are only $\mathrm{D_{even}}$-type (and possibly also A-type) submodels. On an LG variable of weight $w$, the orbifold group generally acts as $x \to e^{2i\pi w/H} x$.

\begin{table}
  \centering
  \caption{LG superpotentials corresponding to ADE-type minimal models}\label{ADE LG superpotentials}
  \begin{tabular}{|c|c|c|}
    \hline
    PF type & LG superpotential & minimal model level\\
    \hline
    $A$ & $x^{k+2}$ & $k\geq1$ \\
    $D$ & $x^\frac{k+2}{2}+xy^2$ & $k \geq 4,~k\in 2\mathbb{Z}$ \\
    $E_6$ & $x^3+y^4$ & $k=10$ \\
    $E_7$ & $x^3+xy^3$ & $k=16$ \\
    $E_8$ & $x^3+y^5$ & $k=28$ \\
    \hline
  \end{tabular}
\end{table}

\subsection{RR ground states of Gepner models}

In the untwisted sector, the A-type level $k$ minimal model has $k+1$ supersymmetric RR ground states $|l\rangle$, labeled by $l=0,1,...,k$. They have R-charge $q=\Tilde{q}=\frac{l+1}{k+2}-\frac{1}{2}$, and can be represented as $x^l$. In the $\nu$-twisted sector, there is a unique supersymmetric RR ground state with R-charge $q=-\bar{q}=\frac{l_\nu+1}{k+2}-\frac{1}{2}$, where $l_\nu \equiv -\nu -1 ~ (\mathrm{mod}~k+2)$. In the case of the $4_D$ minimal model, the charge of a Ramond ground state with given $l$ is unchanged, but the spectrum is different. One can see by expanding \eqref{eq:ZD} that only states with even $l$, namely $l=0,2,4$ are consistent with modular invariance of the partition function. Besides, there are two states with $l=2$ in this model, due to the factor of 2 in the affine modular invariant. 

For all types of modular invariants, the condition for a state to survive the orbifold projection of the theory is that the total R-charge satisfies
\begin{equation}
    \sum_iq_i=\frac{1}{2}+\mathbb{Z}.
    \label{eq:halfZ}
\end{equation}
A subtlety can arise with trivially twisted states, which can appear when $\nu$ is a divisor of $H$. The resulting states will be obtained by the tensor product of the twisted states from the non-trivially twisted submodels and the untwisted states from the trivially twisted submodels.

In the LG/CY correspondence, the Dolbeaut cohomology on both sides identifies states with R-charge $(q,\bar{q})$ with cohomology classes $H^{\frac{3}{2}-q,\frac{3}{2}+\bar{q}}$. We recover the property of a $CY_3$ where $h^{0,0}=h^{3,3}=h^{3,0}=h^{0,3}=1$. Here, the (3,0)-form and the (0,3)-form originate from the untwisted sector while the (0,0)-form and the (3,3)-form lie in the first and last twisted sector, respectively. In the following, we give two examples of the computation of the Hodge number $h^{2,1}$, and the details of the complex structure deformations of the remaining models from the ones listed above is found in \autoref{app:RRgs}.

Although it is perfectly possible to list all RR ground states of Gepner models by finding the $l_i$ solutions to \eqref{eq:halfZ}, we will take a different approach in these examples. On the LG side of the RG flow, the (2,1) (i.e. complex structure) deformations are represented by monomial deformations of the LG superpotential that are invariant under the orbifold group action on the LG variables, and modded by the equations of motion. This can be intuitively understood because the superpotential is related to the algebraic equation defining the Calabi-Yau space.\\

\noindent \textbf{Example: $1^3 \times 4^3_A$ model}\\

\begin{table}[t]
  \centering
  \caption{(2,1)-forms in the $1^3\times4_A^3$ model}\label{1^3-4^3A}
  \begin{tabular}{|c|c|c|}
    \hline
    sector & monomial representations & counting \\
    \hline
    \multirow{12}{*}{untwisted} & $x_1x_2x_3$ & 1 \\
                            & $x_ix_jy_m^2$ & 9 \\
                            & $x_ix_jy_my_n$ & 9 \\
                            & $x_iy_m^4$ & 9 \\
                            & $x_iy_m^3y_n$ & 18 \\
                            & $x_iy_m^2y_n^2$ & 9 \\
                            & $x_iy_m^2y_ny_l$ & 9 \\
                            & $y_m^4y_n^2$ & 6 \\
                            & $y_m^4y_ny_l$ & 3 \\
                            & $y_m^3y_n^3$ & 3 \\
                            & $y_m^3y_n^2y_l$ & 6 \\
                            & $y_1^2y_2^2y_3^2$ & 1 \\
    \hline
        $\nu=3$ twisted & $x_1 x_2 x_3$ & 1 \\
    \hline
  \end{tabular}
\end{table}

\noindent The LG potential for this model is
\begin{equation}
    \mathcal{W}=\sum_{i=1}^3x_i^3+\sum_{m=1}^3y_m^6 + z^2,
\end{equation}
with relative weights $[x]=2$, and $[y]=1$. Note the presence of the mass term $z^2$ because there is an even number of A-type fields, although it does not contribute to the counting of chiral primaries. The model is divided by a $\Z_6$ orbifold action
\begin{equation}
    g : (x_i,y_m,z) \to (e^{2i\pi/2}x_i, e^{i\pi/3}y_m, -z).
    \label{eq:Z6}
\end{equation}
The chiral primaries (corresponding to (2,1)-forms) belong to the chiral ring
\begin{equation}
    \C[x_i,y_m,z]/\langle \partial_{x_i} \W, \partial_{y_m} \W, \partial_z \W \rangle
\end{equation}
which corresponds to the monomials modded out by the equations of motion of the LG variables $\partial \W = 0$, or mathematically, the Jacobian ideal of the polynomial ring. In the chiral ring, the orbifold projection selects the monomials that are invariant under the $\Z_6$ action \eqref{eq:Z6}. This leaves us with weighted degree 6 monomials. There are 83 such chiral primaries in the untwisted sector, and one chiral primary comes from the middle-twisted sector for a total of $h^{2,1} = 84$. The combinations are shown in Table \ref{1^3-4^3A}.

Note how in the $\nu = 3$ twisted sector, the $x_i$ fields become untwisted (or trivially twisted) and the twisted ground state of the level-4 models has R-charge $q = - \bar{q} = 0$. Consequently, the state from this sector appears to have only ``untwisted charge''.\\

\noindent \textbf{Example: $1^3 \times 4^3_D$ model}\\

\begin{table}[t]
  \centering
  \caption{(2,1)-forms in the $1^3\times4^3_D$ model}\label{1^3-4^3D}
  \begin{tabular}{|c|c|c|}
    \hline
    sector & monomial representations & counting \\
    \hline
    \multirow{13}{*}{untwisted} & $x_1x_2x_3$ & 1 \\
                            & $x_ix_jy_m$ & 9 \\
                            & $x_ix_jz_m$ & 9 \\
                            & $x_iy_mz_n$ & 18 \\
                            & $x_iy_m^2$ & 9 \\
                            & $x_iy_my_n$ & 9 \\
                            & $x_iz_mz_n$ & 9 \\
                            & $y_m^2y_n$ & 6 \\
                            & $y_m^2z_n$ & 6 \\
                            & $y_my_nz_l$ & 3 \\
                            & $y_mz_nz_l$ & 3 \\
                            & $y_1y_2y_3$ & 1 \\
                            & $z_1z_2z_3$ & 1 \\
                            
    \hline
  \end{tabular}
\end{table}

\noindent We provide this example for its similarity with the previous one and to illustrate the differences between A and D-type modular invariants at the level of the LG theory. The worldsheet superpotential reads
\begin{equation}
    \mathcal{W} = \sum_{i=1}^3 x_i^3 + \sum_{m=1}^3 (y_m^3 + y_m z_m^2).
\end{equation}
Note that in the absence of $4_A$ field, all LG variables have equal weights and no trivial field is needed since there is an odd number of A-type factors. This is a case where $H=\mathrm{lcm}\{k_i + 2\}/2$, and the orbifold group is then $\Z_3$ with action
\begin{equation}
    g : (x_i, y_m, z_m) \to (e^{2i\pi/3}x_i, e^{2i\pi/3}y_m, e^{2i\pi/3}z_m).
\end{equation}

The equations of motion for the D-type variables give $y_m^2 \sim z_m^2$ and $y_m z_m \sim 0$. The existence of two LG variables for one D-type model is related to the presence of two $l=2$ states in the corresponding minimal model. In this example all the 84 (2,1)-forms are from the untwisted sector, see Table \ref{1^3-4^3D}.

\section{B-branes and matrix factorization}
\label{sec:B-branes_MF}

With the objective of Gepner models string compactifications in mind, we have described the spectrum of Ramond ground states and NS chiral primaries of the CFT, that are in one-to-one correspondence with the supergravity complex structure moduli. Another essential aspect of flux compactification are the D-branes and cycles of the compact space. Those objects can be described from the Gepner CFT or LG QFT point of view. They provide a basis of cycles on which the NSNS and RR fluxes can be turned on, as was extensively studied in \cite{Becker_2007}. As we focus on the tadpole cancellation condition, we will draw our attention to the overlap between D-branes and twisted Ramond ground states, and the O-planes topological classes in terms of D-branes classes.

We follow \cite{Becker_2007} and \cite{Walcher_2005} to express B-type D-branes (or the so-called B-branes) $[\Lambda_n] ~ (n=0,1,...,H-1)$ in the language of matrix factorization (without going deep into the category theory) developed in \cite{Kapustin_2003,Kapustin_2004,Hori_2008,Hori_2004,Walcher_2005,kapustin2003topologicalcorrelatorslandauginzburgmodels}. At the level of the BCFT, A- and B-branes are defined by the boundary conditions on the $\mathcal{N}=(2,2)$ supercurrents in the open string sector
\begin{equation}
    \begin{aligned}
        \text{A-branes: } ~ G_L^{\pm}(z) & = G_R^\mp(\bar{z}) ~ \text{ at } ~ z=\bar{z}\\
        \text{B-branes: } ~ G_L^{\pm}(z) & = G_R^\pm(\bar{z}) ~ \text{ at } ~ z=\bar{z}
    \end{aligned}
\end{equation}
Both A- and B-branes have a description in the LG/Gepner formalism and are of interest for flux compactifications \cite{Becker_2007}. In our case, since we will mainly work in the mirror picture, only B-branes will be needed and we will not further expand the topic of A-branes.

We will compute the intersection matrix of B-branes and read off the B-brane charges, understood as the overlap between B-branes and twisted Ramond ground states. In the end, we will see how to give a minimal basis of the B-branes and explain which B-branes can support fluxes and contribute to the tadpole cancellation condition.

\subsection{Matrix factorization for Landau-Ginzburg orbifolds}

A B-brane in an LG orbifold with $r$ fields is specified by a set of data $(M,\sigma,Q,\rho)$ \cite{Hori_2008}, where $M$ is a graded module with $\Z_2$-grading $\sigma$ over the polynomial ring $\C [x_1,...,x_r]$. $Q$ is a boundary fermionic degrees of freedom that fixes B-type supersymmetry on the open string action. It is represented by a matrix factorization of the LG superpotential $\W(x_1,...,x_r)$, which is an odd endomorphism on $M$
\begin{equation}
    Q(x)= \begin{pmatrix} 0 & f(x) \\ g(x) & 0 \end{pmatrix},
\end{equation}
such that
\begin{equation}
    f(x) \cdot g(x) = g(x) \cdot f(x) = \W(x) \cdot \mathbb{1}_r ~ \text{ or equivalently } Q(x)^2 = \W(x) \cdot \mathbb{1}_{2r}~.
\end{equation}
$\rho$ is an even representation of the orbifold group $\Gamma$. The invariance under $\Gamma$-actions is given on the matrix factorization by the condition
\begin{equation}
    \rho(g)Q(gx)\rho(g)^{-1}=Q(x),~~~~~~\forall g\in\Gamma,
    \label{eq:QZinvariance}
\end{equation}

Note that the open string spectrum between two branes $(M_1, \sigma_1, Q_1, \rho_1)$, $(M_2, \sigma_2, Q_2, \rho_2)$ is the space of homomorphisms of the modules $\mathrm{Hom}(M_1, M_2)$. Matrix factorizations of a given LG superpotential $\W$ form a differential category $\mathcal{MF(W)}$. We will neither discuss the open string states, nor the categorical description of matrix factorizations in the following. We refer the reader to \cite{Walcher_2005, Hori_2008, Kapustin:2003ga} for further details.

\subsection{Applications to Gepner models}

\label{sec:Gepner_branes}

We first give the matrix factorization for a single minimal model and the relevant $\mathbb{Z}_H$ representation. For A-type and D-type minimal models, the matrix factorizations are respectively given by
\begin{equation}
    Q=\begin{pmatrix}
0 & x \\
x^{k+1} & 0
\end{pmatrix},~~~\text{for A-type}~~~~~~\text{and}~~~~Q=\begin{pmatrix}
0 & x \\
x^\frac{k}{2}+y^2 & 0
\end{pmatrix},~~~\text{for D-type}.
\label{eq:Qsm}
\end{equation}
The $\mathbb{Z}_H$ action on $Q$ is represented by 
\begin{equation}
    \rho(g)=\omega^n
\begin{pmatrix}
1 & 0 \\
0 & \omega
\end{pmatrix},~~\forall g\in\Gamma,
    \label{eq:Zrep}
\end{equation}
where $\omega=\mathrm{e}^\frac{2\pi i}{k+2}$ and $n=0,1,...,H-1$. Note that we can in principle choose a different matrix factorization together with a proper orbifold group representation, which may or may not change the basis of B-branes depending on the particular models. In this paper, we will use this way of matrix factorization to get a basis of B-branes and then express internal D0-branes and O-planes in terms of this basis, as long as only the brane charges are concerned.
If there happens to be a trivial minimal model (which we may drop from time to time), the only matrix factorization is
\begin{equation}
    Q=
\begin{pmatrix}
0 & x \\
x & 0
\end{pmatrix},
\end{equation}
with the $\mathbb{Z}_2$ generator represented by 
\begin{equation}
    \rho(g)=\pm
\begin{pmatrix}
1 & 0 \\
0 & -1
\end{pmatrix}.
\end{equation}

Now we are ready to consider matrix factorization for the Gepner models which are orbifolded tensor products of minimal models. The matrix factorization for the tensor product theory reads
\begin{equation}
    Q=\sum_{i=1}^r \sigma_1 \otimes ... \otimes \sigma_{i-1} \otimes Q_i \otimes \mathbb{1} \otimes ... \otimes \mathbb{1},
    \label{eq:Qtensor}
\end{equation}
where $r$ is the number of minimal model factors as before, $\otimes$ denotes the ordinary Kronecker product, $\mathbb{1}$ denotes the identity $2\times 2$ matrix, $\sigma = \mathrm{diag}(1,-1)$ is the grading of the two-dimensional module on which a single $Q_i$ acts, and the subscript $i$ labels the $i^\mathrm{th}$ minimal model factor. Having factors of $\sigma$ on the left of $Q_i$ and identity matrices on the right ensures the graded structure of the tensor product. It follows from \eqref{eq:Qtensor} that $Q^2 = \W \cdot \mathbb{1}$. The grading of the full module $M_r$ on which $Q$ acts is $\sigma_r = \sigma^{\otimes r}$, which shows that $Q$ is an odd endomorphism on $M_r$. A representation of the $\mathbb{Z}_H$ orbifold action on $M_r$ is now
\begin{equation}
    \rho(g)=\rho_1(g_1)\otimes\rho_2(g_2)...\otimes\rho_r(g_r),
\end{equation}
where the subscript again labels the minimal model factor. Firstly, it's easy to check that $\rho(g)Q(gx)\rho(g)^{-1}=Q(x)$ is satisfied. Secondly, one can notice
that for a given $g\in\Gamma$, only the number $n=\sum_{i=1}^r\frac{n_iH}{k_i+2} ~ \mathrm{mod}~H$ matters, so we will use it to label the B-branes $\L_n$ and the corresponding $g_{(n)}$.

In a given twisted sector $\nu=1,2,...,H-1$, there may be some terms of the superpotential that are left invariant by the action of the orbifold group element. The corresponding fields remain untwisted, or trivially twisted. There are $r_\nu$ such fields per twisted sector, that we will call $x_i^u$, while the twisted fields are called $x_i^t$. The B-brane charge in an orbifolded LG-model can be computed using the following equation\footnote{We may use $g$ and its matrix representation $\rho(g)$ interchangeably when there is no confusion. Also, $Q_{(\nu)}$ shall not be confused with $Q_i$(the matrix factorization for the $i^\mathrm{th}$ minimal model). The former is the truncated one in the tensor product model and depends on which twisted sector one is focusing on.}\cite{Walcher_2005},
\begin{equation}
    \langle\L_n|\nu;\alpha\rangle=\frac{1}{r_\nu!}\oint\frac{\phi_\nu^\alpha\mathrm{Str}[g_{(n)}^\nu\partial Q_{(\nu)}^{\wedge r_\nu}]}{\partial_1\mathcal{W}_\nu...\partial_{r_\nu}\mathcal{W}_\nu},\label{B-brane charge}
\end{equation}
where the quantities that appear in \eqref{B-brane charge} are defined by turning the twisted fields to 0, more specifically, define $\W_\nu = \W(x_i^u,x_i^t=0)$ and $Q_{(\nu)}=Q(x_i^u,x_i^t=0)$. The term $\partial Q_{(\nu)}$ is a matrix-valued one-form with explicit expression
\begin{equation}
    \partial Q_{(\nu)}=\sum_{i=1}^{r_\nu}\partial_{x_i^u}Q_{(\nu)}dx_i^u.
\end{equation}
The term $\partial Q_{(\nu)}^{\wedge r_\nu}$ means we are taking the canonical $r_\nu$-fold wedge product. The supertrace is simply given by $\mathrm{Str}(Q) = \mathrm{Tr}(\sigma_{r_\nu} Q)$, with the grading represented by $\sigma_{r_\nu} = \mathrm{diag}(1,-1)^{\otimes r_\nu}$. The untwisted fields act on the lowest energy RR state as chiral primaries, with $q_L^u=q_R^u=\hat{c}^u/2$, where $\hat{c}^u/2$ is the central charge corresponding to $\mathcal{W}_\nu$. They are described by $\phi_\nu^\alpha$ in \eqref{B-brane charge} with the expression
\begin{equation}
    \phi_\nu^\alpha=\prod_{i=1}^{r_\nu}(x_i^u)^{\alpha_i},
\end{equation}
where $\alpha_i$ are integers such that $\phi_\nu^\alpha$ has the expected RR charge. Let us note that when there are no trivially twisted states ($r_\nu=0$), \eqref{B-brane charge} reduces to the usual $\mathrm{Str}(g_{(n)}^\nu)$ as has been used, for example, in \cite{Becker_2007}.

The overlaps between the B-branes and the twisted Ramond ground states give us access to the intersection matrix of the B-branes $[\L_n]$ through the index theorem \cite{Walcher_2005}
\begin{equation}
    \Tr_{\mathcal{H}_{\L_m,\L_n}}(-1)^F = \frac{1}{H} \sum_{\nu = 1}^{H-1} \sum_{\alpha, \beta} \langle \L_m | \nu ; \alpha \rangle \frac{\eta_\nu^{\alpha \beta}}{\prod_{\nu q_i \notin \mathbb{Z}} (1-\omega_i^\nu)} \langle \nu ; \beta | \L_n \rangle~,
    \label{eq:Bbintmat}
\end{equation}
where $q_i=[X_i]/H$ is the weight of the LG-variable and $\eta_\nu^{\alpha \beta}$ is the inverse metric of
\begin{equation}
    \eta_{\alpha \beta}^\nu = \mathrm{Res}_\nu (\phi_\nu^\alpha \phi_\nu^\beta)~.
\end{equation}

\noindent \textbf{Example: $1^3 \times 4^3_A \times 0 $ model}\\

\noindent The superpotential in this model reads
\begin{equation}
    \mathcal{W} = x_1^3 + x_2^3 + x_3^3 + y_4^6 + y_5^6 + y_6^6 + z^2.
    \label{eq:1^3x4^3W}
\end{equation}
Note that in we restore the trivial minimal model here as it is necessary to obtain the correct charges of B-branes. The matrix factorization and the corresponding $\mathbb{Z}_6$ action are represented in each minimal model by(with $\omega=\mathrm{e}^\frac{i\pi}{3}$)
\begin{equation}
    Q_{i=1,2,3}=\begin{pmatrix}
0 & x_i \\
x_i^2 & 0
\end{pmatrix},~~~
Q_{i=4,5,6}=\begin{pmatrix}
0 & y_i \\
y_i^5 & 0
\end{pmatrix},~~~
Q_{z}=\begin{pmatrix}
0 & z \\
z & 0
\end{pmatrix}.
\end{equation}
The $\nu=2$ and $\nu=4$ sectors are not populated in the $\mathbb{Z}_6$ orbifolded model (there are no RR ground states in these sectors), while the $\nu=1$ and $\nu=5$ sectors are purely twisted. In the $\nu=3$ sector, the fields $x_1,~x_2,~x_3$ are untwisted and do contribute to $(2,1)$-forms as well as B-brane charges in the semi-classical analysis. We will then focus on this sector in the following. In this sector, the truncated worldsheet superpotential becomes
\begin{equation}
    \mathcal{W}_l=x_1^3+x_2^3+x_3^3,
\end{equation}
so we have $r_\nu=3$, $\hat{c}^u=3$ and the only $\phi$ to be considered here is $\phi_3^1=x_1x_2x_3$. The matrix factorization in the effective (restricted) theory reads
\begin{equation}
Q_{(3)}= Q_1 \otimes \mathbb{1} \otimes \mathbb{1} \otimes \mathbb{1} \otimes \mathbb{1} \otimes \mathbb{1} \otimes \mathbb{1} + \sigma \otimes Q_2 \otimes \mathbb{1} \otimes \mathbb{1} \otimes \mathbb{1} \otimes \mathbb{1} \otimes \mathbb{1} + \sigma \otimes \sigma \otimes Q_3 \otimes \mathbb{1} \otimes \mathbb{1} \otimes \mathbb{1} \otimes \mathbb{1}.
\end{equation}
We compute
\begin{equation}
    \partial Q_{(3)}^{\wedge3}=6(\partial_1Q_1\otimes\partial_2Q_2\otimes\partial_3Q_3\otimes\mathbb{1}\otimes\mathbb{1}\otimes\mathbb{1})dx^1\wedge dx^2\wedge dx^3,
\end{equation}
and hence
\begin{equation}
    \mathrm{Str}[g_{(n)}^3(Q_{(3)})^{\wedge3}] \propto \Tr \begin{pmatrix} 0 & 1 \\ 2x_i & 0 \end{pmatrix}=0,
\end{equation}
which is independent of $n$. For $\nu=1$ or $\nu=5$, one can simply use $\mathrm{Str}(g_{(n)}^\nu)$ to compute the overlap, which we will not show explicitly here. We use \eqref{eq:Bbintmat} to compute the intersection matrix of B-branes of this model and we find
\begin{equation}
    \Tr (-1)^F = \begin{pmatrix} 0 & -3 & -3 & 0 & 3 & 3 \\ 3 & 0 & -3 & -3 & 0 & 3 \\ 3 & 3 & 0 & -3 & -3 & 0 \\ 0 & 3 & 3 & 0 & -3 & -3 \\ -3 & 0 & 3 & 3 & 0 & -3 \\ -3 & -3 & 0 & 3 & 3 & 0 \end{pmatrix}~.
\end{equation}

Firstly, we notice that this matrix has rank 2, so that the B-brane basis has two linearly independent elements. In that case $[\L_{n+3}] = - [\L_n]$\footnote{The bracket denotes the homological classes, which we sometimes drop when there is no confusion.}, so we will in general choose two branes out of $\L_0, \L_1, \L_2$. Secondly, restricting $\Tr (-1)^F$ to e.g. $\L_0, \L_1$, it does not have determinant 1. This means that the $\L_n$'s do not contain a minimal basis of the B-branes charges, and such a basis will prove necessary in the computation of the tadpole cancellation condition. We can find a minimal basis of branes by constructing permutation branes. These branes are described by a matrix factorization which is not obtained as the naive tensor product of each minimal model\cite{Brunner_2005}. In our example, the piece of the superpotential $x_1^3 + x_2^3$ can be factorized by the matrix
\begin{equation}
    Q_{(12)} = \begin{pmatrix} 0 & x_1 + x_2 \\ x_1^2 - x_1 x_2 + x_2^2 & 0 \end{pmatrix}~.
    \label{eq:perm_branes}
\end{equation}
This matrix factorization still satisfies the $\Z_6$-invariance \eqref{eq:QZinvariance} with $\Z_6$ representation \eqref{eq:Zrep}. This means that the matrix factorization for the tensor product theory reads
\begin{equation}
    Q = Q_{(12)} \otimes \underbrace{\mathbb{1} \otimes ... \otimes \mathbb{1}}_{\text{5 times}} + \sum_{i=2}^6 \sigma \otimes ... \otimes \underbrace{Q_i}_{\text{position }i} \otimes ... \otimes \mathbb{1}
\end{equation}
and the corresponding $\Z_6$ representation is
\begin{equation}
    \rho(g) = \rho(g)_{\omega=e^{2i\pi/3}}^{\otimes 2} \otimes \rho(g)_{\omega=e^{i\pi/3}}^{\otimes 3} \otimes \rho(g)_{\omega=-1}
\end{equation}
Note there are now 6 tensored matrices instead of 7, so that the matrices $Q$ and $\rho(g)$ are smaller than before. The branes corresponding to this factorization are written $\L_n^{(12)}$, and we can still compute their overlap with twisted Ramond ground states using \eqref{B-brane charge}. Their charge class satisfies
\begin{equation}
    [\Lambda_n^{(12)}] = \frac{[\L_n] + [\L_{n+1}]}{3}~.
\end{equation}

For some model it may be necessary to build permutation branes with more than one ``permutation'' but in the case at hand, the basis given by the branes $\L_0^{(12)}, \L_1^{(12)}$ yields the intersection matrix
\begin{equation}
    \Tr (-1)^F = \begin{pmatrix} 0 & -1 \\ 1 & 0 \end{pmatrix}~.
\end{equation}

One may also consider adding two extra trivial squared terms to the superpotential \eqref{eq:1^3x4^3W}, which leads to a different geometric formulation (see \ref{sec:branes_dim}). In that case, these two extra factors can be tensored as a permutation brane \cite{Becker_2007}, leaving the brane basis and its intersection matrix unchanged. Adding two more trivial minimal models, however, changes the O-plane classes as will be seen in \autoref{sec:orientifolds_gepmod}, because the Kn\"orrer periodicity is doubled compared with the ordinary case where there is no orientifold \cite{Walcher_2005}.

\subsection{Geometric interpretation of B-branes charges}

\label{sec:branes_dim}

We were able to compute the charges of B-branes in the CFT language, however an essential condition for type IIB moduli stabilization is the D3-brane tadpole cancellation condition \eqref{eq:tadcond} (see \autoref{sec:fluxquant} for further details). Since this condition is geometric, we need to understand the large-volume geometric interpretation of the BCFT boundary states. In particular, we need to identify the point class of the corresponding geometry, as explained below. We will compute those classes in the rigid Calabi-Yau manifold which is the mirror picture of the ``non-geometric" LG model. Since the tadpole cancellation condition is topological, it should not depend on the position of the theory in the K{\"a}hler moduli space.

Suppose one unit of RR flux and NSNS flux are supported on a 3-cycle $A$ and another 3-cycle $B$, then the tadpole contribution from the fluxes is \cite{Becker_2007}
\begin{equation}
    [\mathrm{Flux}]=(A\cap B)[\mathrm{pt.}],
\end{equation}
where $[\mathrm{pt.}]$ is the class of a point where a D3-brane can sit on. From the internal viewpoint, they correspond to D0-branes (since we are interested here in 4D space-time filling branes). This implies that one has to figure out the dimension of the D-branes (represented before by matrix factorization) in the large volume regime. Under the CY/LG correspondence, the branes $[\Lambda_0],...,[\Lambda_{n-1}]$ can be categorically described by the exceptional collection $\{E_0,E_1,E_2,...,E_{n-1}\}$, which are the dual exceptional collection of line bundles $\{\mathcal{O},\mathcal{O}(1),\mathcal{O}(2),...,\mathcal{O}(n-1)\}$ of the ambient space, restricted to the elliptic curve/surface \cite{Douglas_2001,ASPINWALL_2005}. Here, ``dual'' is to be understood with respect to the Euler pairing, i.e.
\begin{equation}
    \chi(E_i,O(j))=\delta_{ij},
\end{equation}
where $\chi$ is the Euler pairing and can be computed using the Hirzebruch-Riemann-Roch formula,
\begin{equation}
    \chi(E_i,E_j)=\int_M\mathrm{ch}(E_i^\vee)\mathrm{ch}(E_j)\mathrm{Td}(T_M),\label{eq:eulerpairing}
\end{equation}
and $E_i^\vee$ is the dual bundle of $E_i$. For the (weighted) projective space, a more useful formula to compute the Euler pairing of line bundles is
\begin{equation}
     \chi(\mathcal{O}(i),\mathcal{O}(j))=\chi(\mathcal{O}(j-i)),~~i\leq j,
\end{equation}
where $\chi(\mathcal{O}(i-j))$ counts the number of holomorphic sections of $\mathcal{O}(i-j)$, which is equal to the number of (weighted) degree $i-j$ in $(\mathbb{W})\CP^d$. The ``dimension'' of the B-branes (in terms of bound states of D-branes) can be read off from the Chern characters of these bundles, since in projective space they are expanded as
\begin{equation}
    \mathrm{ch}(E) = c_0 + c_1 H + ... + c_{d-1} H^{d-1}
\end{equation}
where $H$ is a complex 1-form, Poincaré dual of the hyperplane class of projective space. Consequently, a pure D0-brane (which is a codimension $d$ object) can only have a non-vanishing $(d-1)^{\text{th}}$ Chern character. Note that if the hypersurface we are interested in intersects the divisor of projective space at $n$ distinct points, the Chern character of a D0-brane follows $\mathrm{ch(D0)} = (0 , ... , 0 , 1/n)^{\mathrm{T}}$ on the hypersurface. If none of the $[\Lambda_i]$'s corresponds to a D0-brane, one should consider other basis of branes. An option that proved useful are the permutation branes previously introduced. Below we take the $1^3 \times 4^3_A$ model as an example and follow \cite{Becker_2007}, with more details, to explain the above arguments.\\

\noindent \textbf{Example: $1^3 \times 4^3_A$ model}\\

\noindent The idea in this LG/CY correspondence is that one can consider the chiral fields in the LG model as some ambient space (homogeneous) coordinates and perform the change of variables when doing path integral. Integrating w.r.t. the new variables that are linear gives rise to a delta function (or a product of delta functions) which essentially defines a hypersurface in the ambient space. The single valuedness of this change of variable requires the hypersurface to be at some special symmetric point of the moduli space. To get the correct spacetime spectrum and topological invariants such as Euler characteristic, one has to perform a further GSO-like projection, which usually takes the form of a quotient by $\mathbb{Z}_H$ or $\mathbb{Z}_H\times\mathbb{Z}_H$ depending on the factorization of the full $c=9$ theory.

As has been mentioned in the introduction, there is an ambiguity in the geometric description of the $1^3 \times 4^3_A$ model that we will address later on. It can be described as the mirror manifold of a rigid Calabi-Yau orbifold, either $(T^2)^3/\Z_3 \times \Z_3$ or $T^2 \times K3/\Z_6$. 

For the time being, let us focus on $T^2 \times K3 / \Z_6$ as it includes the simple case of the $1^3$ model description of $T^2 : \{(x,y,z) \in \CP^2 ~ ; ~ x^3 + y^3 + z^3 = 0\}$  at the $\mathbb{Z}_3$ symmetric point of both the complex structure and the K\"ahler moduli spaces\footnote{$T^2$ has a complex structure $\tau=\mathrm{e}^\frac{2\pi i}{3}$ and the generators of $\mathbb{Z}_3\times\mathbb{Z}_3$ act on the holomorphic coordinates of $T^6$ as $\tilde{g_{12}}:(z_1,z_2,z_3)\rightarrow(\omega z_1,\omega^{-1}z_2,z_3)$, $\tilde{g_{23}}:(z_1,z_2,z_3)\rightarrow(z_1,\omega z_2,\omega^{-1}z_3)$, where $\omega$ is the third root of unity.}. The ambient space for this model is $\mathbb{CP}^2$, which has a full exceptional collection
\begin{equation}
    \{\mathcal{O},\mathcal{O}(1),\mathcal{O}(2)\}.\label{line bundle excp.}
\end{equation}
The Chern characters of these line bundles are easily expressed in terms of $H$ as
\begin{equation}
    \mathrm{ch}(\mathcal{O}(n))=\mathrm{e}^{nH},
\end{equation}
where we temporarily neglect the fact that the j-th Chern character($j\geq2$) vanishes for a complex line bundle, since we may need such terms when computing the Euler pairing. The Todd class of the tangent bundle of $\mathbb{CP}^2$ is computed to be
\begin{equation}
    \mathrm{Td}(T\mathbb{CP}^2) = \qty(\frac{H}{1-\mathrm{e}^{-H}})^3.
\end{equation}
The Euler pairing \eqref{eq:eulerpairing} expressed in a matrix form is
\begin{equation}
    S_{ij}=\chi(\mathcal{O}(i),\mathcal{O}(j))=
\begin{pmatrix}
1 & 3 & 6 \\
0 & 1 & 3 \\
0 & 0 & 1
\end{pmatrix}.
\end{equation}
The vector bundles \{$E_n$\} associated with $[\Lambda_n]$ in fact form the dual exceptional collection of \eqref{line bundle excp.}, in the sense that
\begin{equation}
    \chi(E_n,\mathcal{O}(m))=\delta_{nm}.
\end{equation}
Their Chern characters are easily computed to be
\begin{equation}
    \mathrm{ch(E_n)} = (S^{-1})_\mathrm{nm}\mathrm{ch(\mathcal{O}(m))},
\end{equation}
which leads to (3.22) in \cite{Becker_2007}, given below as well.
\begin{equation}
    \mathrm{ch}_i(\Lambda_n)=\begin{pmatrix}
1 & -2 & 1 \\
0 & 1 & -1
\end{pmatrix}.
\end{equation}
Obviously, none of these $[\Lambda_n]$'s is point-like (they all bear D2 charges) and thus they cannot be considered as D0-branes. We can use the adjunction formula to implement the restriction. Denoting the torus by $Y$, we have
\begin{equation}
    \mathrm{Td}(T Y)=\frac{\mathrm{Td}(T\mathbb{CP}^2)}{\mathrm{Td}(NY_{\mathbb{CP}^2})},
\end{equation}
where $N Y_\mathbb{CP}^2$ is the normal bundle of $Y$ in $\mathbb{CP}^2$, whose Todd class is
\begin{equation}
     \mathrm{Td}(N Y_{\mathbb{CP}^2})=\frac{3H}{1-\mathrm{e}^{-3H}}.
\end{equation}
Now we can compute the Euler pairing matrix of $\wedge^n\Omega(n)|_Y$, which is interpreted as the intersection matrix of $[\Lambda_n]$'s.
\begin{equation}
    \mathbf{I}_{mn}=\begin{pmatrix}
0 & 3 \\
-3 & 0
\end{pmatrix},
\end{equation}
where $m,n$ run from 0 to 1, since we know that $[\Lambda_2]$ is not independent of the others due to the $\mathbb{Z}_3$ symmetry. From this computation, we see that $[\Lambda_0]$ and $[\Lambda_1]$ do not give a minimal basis of B-branes in the $1^3$ model. In the LG limit (at the Gepner point), we note that the D-brane charges of $\Lambda_n$ are \cite{Becker_2007}
\begin{equation}
    \braket{\Lambda_n}{k} = \mathrm{Str}(g_{(n)}^k)=(1-\omega^{2k})^3\omega^{2kn},
\end{equation}
where $\ket{k=1,2}$ are the Ramond ground states in the two twisted sectors, respectively, and $\omega = e^{i\pi/3}$ here. The intersection matrix of $\Lambda_n$'s $(n=0,1)$ is\footnote{Apply Eqn.(2.35) in \cite{Becker_2007} or Eqn.(5.14) in \cite{Walcher_2005} for the $1^3$ model.}
\begin{equation}
    \mathbf{I}_{mn}=\mathrm{Tr}(-1)^F=\begin{pmatrix}
0 & 3 \\
-3 & 0
\end{pmatrix},
\end{equation}
which agrees with the large volume computation. Now we need the permutation branes $[\L^{(12)}_n]$ introduced in \eqref{eq:perm_branes}. They form a minimal basis of branes and it is now straightforward to see which brane corresponds to a D0-brane by computing the Chern characters. By comparing $\langle\L^{(12)}_n|k\rangle$ and $\langle\Lambda_n|k\rangle$, one can see that
\begin{equation}
    [\L^{(12)}_m]=P_{mn} [\Lambda_n],
\end{equation}
where $n=0,1$ and the change of basis matrix is $P=\begin{pmatrix}
\frac{2}{3} & \frac{1}{3} \\
-\frac{1}{3} & \frac{1}{3}
\end{pmatrix}$. Therefore, the Chern characters of the permutation branes are
\begin{equation}
    \mathrm{ch}_i(\L^{(12)}_n)=\begin{pmatrix}
0 & -1 \\
\frac{1}{3} & \frac{1}{3}
\end{pmatrix},
\end{equation}
which implies that $[\L^{(12)}_0]$ corresponds to the point class, namely, a D0-brane in the internal space. Note that the number of D0-branes should correspond to 3 times $\mathrm{ch}_1(\L^{(12)}_0)$, where the factor of 3 comes from the fact that $H$ of $\mathbb{CP}^2$ intersects the elliptic curve at three points \cite{Becker_2007}. It follows that 
\begin{equation}
    [\mathrm{pt.}]_{T^2}=[\L^{(12)}_0] = \frac{2 [\Lambda_0] + [\Lambda_1]}{3},
\end{equation}
so that
\begin{equation}
    \langle [\mathrm{pt.}]_{T^2} | k \rangle = 2(1-\omega^{2k}).
\end{equation}
In a very similar way, one can compute the point class of the $4^3_A$ K3 in $\WCP^3[1,1,1,3]$
\begin{equation}
    \langle [\mathrm{pt.}]_{K3} | k \rangle = 2(1-\omega^k).
\end{equation}
The point class on $T^2 \times K3$ simply is the product of $[\mathrm{pt.}]_{T^2}$ and $[\mathrm{pt.}]_{K3}$. The $\Z_6$ orbifold projection restricts these overlaps to identically charged  Ramond ground states of the $1^3$ and $4_A^3$ CFTs. It also leads to a point class that is invariant under the monodromies of the B-branes, so that
\begin{equation}
    \langle [\mathrm{pt.}]_{T^2 \times K3} | k \rangle = \pm 2(1-\omega^k)^2 (1-\omega^{2k})^2,
\end{equation}
which further implies that
\begin{equation}
    [\mathrm{pt.}]_{T^2 \times K3} = \pm \frac{[\Lambda_1]+[\Lambda_2]}{3}.
\end{equation}

As mentioned earlier, there exists another valid geometric interpretation of the $1^3 \times 4_A^3$ Gepner model as a factorization of three $c=3$ pieces, already noticed in \cite{Fuchs:1989yv}. As explained at the end of \autoref{sec:Gepner_branes}, the superpotential \eqref{eq:1^3x4^3W} can be modified with two additional massive LG fields (equivalently trivial superconformal minimal models). This keeps an odd number of A-type minimal models and does not change the physics of the theory (before orientifold projection). This new superpotential can be split into three identical $T^2$'s of defining equation
\begin{equation}
    T^2 : \{(x,y,z) \in \WCP^2[2,1,3] ~ ; ~ x^3 + y^6 + z^2 = 0\}.
\end{equation}
This leads to a different point class than the $T^2 \times K3$ interpretation, namely
\begin{equation}
    [\mathrm{pt.}]_{(T^2)^3} = \pm [\Lambda_0].
\end{equation}

\section{B-type Orientifolds of Gepner models}
\label{sec:orientifolds_gepmod}

We wish to compactify type IIB string theory on an orientifold of the compact space defined by the LG orbifold. The orientifold projection is defined as dividing out the theory by the B-parity of the Gepner model. From the phenomenological point of view it leads to a four dimensional $\mathcal{N}=1$ theory and introduces O-planes. They are defined the locus of the orientifold projection and are negatively charged objects that are necessary to satisfy the tadpole cancellation condition \eqref{eq:tadcond} (see \autoref{sec:fluxquant} for more details). 

\subsection{B-parities at the Gepner point}

A B-parity of the Gepner model is defined as a composition of $\Omega_B$ and $\tau_{m,\sigma}$, where the worldsheet parity $\Omega_B$ maps the superspace coordinates as $\Omega_B(x^\pm,\theta^\pm,\Bar{\theta^\pm})=(x^\mp,\theta^\mp,\Bar{\theta^\mp})$, and $\tau_{m,\sigma}$ is an involution, up to an orbifold group action, on the LG-variables associated with A-type (or sometimes called A-series) minimal models (denoted by $x$ in the following, unless otherwise specified)
\begin{equation}
    \tau_{\mathbf{m},\sigma} : ~ x_i \to \mathrm{e}^{\frac{2\pi im_i}{k_i+2}} \mathrm{e}^{\frac{\pi i}{k_i+2}} x_{\sigma_i} ~~~~(\text{A-type}).
    \label{involution on x}
\end{equation}
In this equation, $m_i$ runs from $0$ to $k_i+1$ and $\sigma$ is an order-two permutation. It is noted that only switch between two variables with the same degree is allowed in order for the R-charge to be invariant, and that involution requires $m_i+m_{\sigma_i}=0, \mathrm{mod}(k_i+2)$. The extra phase $\mathrm{e}^\frac{\pi i}{k_i+2}$ arises as a gauge transformation from the perspective of the gauged linear sigma model (GLSM) associated with the LG-model \cite{Brunner_2007} and takes care of the sign change in the worldsheet superpotential. Note that for D-type minimal models, there is no such GLSM origin and the extra phase should be chosen to be $\mathrm{e}^\frac{2\pi i}{k_i+2}$ to satisfy $\mathcal{W}(\tau(x))=-\mathcal{W}$
\begin{equation}
    \tau_{\mathbf{m},\sigma} : ~ y_i \to \mathrm{e}^{\frac{2\pi im_i}{k_i+2}} \mathrm{e}^{\frac{2\pi i}{k_i+2}} y_{\sigma_i} ~~~~(\text{D-type}),
    \label{involution on y}
\end{equation}
where we use $y$ to formally represent any D-type minimal model variable. As was explained in \cite{Brunner_2007}, a non-trivial quantum symmetry $\Gamma\cong\mathbb{Z}_{H}$ exists for even $H$. For B-type parity, there is one unique non-trivial involutive quantum symmetry dressing by $g_\omega,~\omega=\mathrm{e}^\frac{2\pi i}{H}$, which maps a state in the $\nu$-twisted sector by multiplying a phase $\omega^\nu$. In particular, it acts trivially on the untwisted sector, while by a minus sign on the ``middle-twisted'' sector.

One has to be careful when determining the action of parity on ground states from twisted sectors. In the geometric case, here meaning models with non-vanishing K\"ahler moduli, one has to understand the structure of the K\"ahler moduli space, and find out which parity symmetry corresponds to the identity of the Calabi-Yau manifold in the large volume regime. Since other parities can be obtained from it by dressing global and/or quantum symmetry, one can know the action of all parities on the states at the Gepner point. One such example is the two-parameter model shown in \cite{Brunner_2004,Brunner_2007}. 

In the non-geometric case, since there is no K\"ahler modulus, it seems reasonable to allow the freedom of phases on twisted sectors under quantum symmetry action, which corresponds to a specific choice of discrete torsion of the system \cite{Vafa:1986wx,Brunner_2004_RCFT}. In this paper, we choose the identity on middle twisted sector to be trivial involution dressed with quantum symmetry (hence trivial involution without quantum symmetry dressing projects out all twisted RR ground states). This choice does not affect the string vacua or the moduli stabilization condition as long as one turns off all the twisted flux.

\subsection{O-planes and their charges}

We are now at a stage to classify all the B-type orientifolds and compute their O-plane charges for the models listed in \autoref{sec:review_Gepner}. We mainly follow the ideas and most conventions of \cite{Becker_2007, Brunner_2007, Hori_2008}. With the objective of model building and moduli stabilization in mind, the goal of this section is to derive the D3-brane charge of the orientifold planes of a given LG model as it appears in the D3 tadpole cancellation condition of type IIB string theory. We first compute the charge of the O-planes in terms of B-branes at the CFT level and read off its tadpole charge following the procedure described in the previous section.

The charge class of the O-plane corresponding to a given orientifold projection $\tau$ corresponds to the class given by the overlap between the associated crosscap state $\ket{C_\tau}$ and the Ramond ground state $\ket{\nu ; \alpha}$. These charges are expressed in the same basis as the B-branes charges. This formula was derived in \cite{Hori_2008}
\begin{equation}
    \braket{C_\tau}{\nu ; \alpha} = \sum_{g^l \tau, (g^l \tau)^2 = g^\nu} \chi(g^\nu) c(g^l \tau) ~ \mathrm{Res}_{W_\nu} (\phi_\nu^\alpha C_{g^\nu \tau}) ~ \prod_{i} (1 + (g^l \tau)_i)
\end{equation}
meaning we are summing on the elements of the orientifold group $(g^l \tau)_{0 \leq l < H}$ that square to the orbifold group element $g^\nu$ for the Ramond ground state $\ket{\nu}$. $(g^l \tau)_i$ are the eigenvalues of these orientifold group elements and $\chi(g^\nu) c(\tau)$ is a phase corresponding to quantum symmetry dressing. They verify
\begin{equation}
    c(\tau)^2 \chi(\tau^2) = 1
\end{equation}
and
\begin{equation}
    c(g\tau) = \chi(g)^{-1}c(\tau)~.
\end{equation}
No dressing by quantum symmetry corresponds to $\chi\equiv1$\footnote{There will always be a sign ambiguity of $c$, which we will choose such that the O-plane charge in (\ref{tadpole cancellation}) is negative, which implies that these are $O^-$ planes implementing $SO(N)$ gauge symmetry.}, while the non-trivial dressing is given by a discrete-torsion-like phase $\chi(g^\nu) = \mathrm{e}^\frac{2\nu\pi i}{H}$. Given this formula, it is possible to compute the O-planes charge classes in terms of the B-brane charge classes.\\

\noindent \textbf{Example: $1^3 \times 4^3_A$ model}\\

\noindent Keeping in mind the $\mathbb{S}_3$ permutations between the 1 and $4_A$ models respectively, there are 3 non-equivalent permutations to compose the involution $\tau$ for this model. They are denoted as $\sigma_0=\mathrm{id}$, $\sigma_1=(12)$, $\sigma_2=(45)$, and $\sigma_3=(12)(45)$. Additionally, a non-trivial phase $\mathbf{m}=(0,0,0,0,0,3)$ (meaning non-equivalent to an orbifold group action) exists for $\sigma_2$ and $\sigma_3$. Since $H=6$ is even, there exists a unique non-trivial involutive dressing by quantum symmetry, which is the dressing by $g_\omega,~\omega=\mathrm{e}^\frac{2\pi i}{6}$ (we denote these by $\Tilde{\sigma}$). It is understood in \cite{Brunner_2007} that quantum symmetry acts trivially on the untwisted sector, however, at the Gepner point, if $n$ twisted ground states survive an orientifold projection with quantum symmetry, then the remaining twisted ground states will survive in cases without quantum symmetry dressing.

Orientifolds for this model are given in Table \ref{1^3-4^3A-0} and \ref{1^3-4^3A-0^3}, corresponding to adding one or three $k=0$ minimal models, and thus corresponding to two different geometries. As noted in Appendix A of \cite{Brunner_2004}, these ``trivial'' models have no influence on the Gepner model itself (in particular on its RR ground states spectrum and on the A- and B-branes), but it has a non-trivial effect on the orientifold and on the O-planes charges. Note that when the model contains three trivial models, the permutations $\sigma$ can also include a permutation between two trivial fields. Those permutations give the same set of branes up to a sign, i.e. relabelling of the twisted Ramond ground states. We recall that the model with three trivial minimal models is the mirror to $T^6/\mathbb{Z}_6\times\mathbb{Z}_6$, and the D0-brane class (or point class) is
\begin{equation}
    [\mathrm{pt.}]_{T^6}= \pm [\Lambda_0]= \pm ([\Lambda_1]-[\Lambda_2]).
\end{equation}
For clarity, we will only consider the choices that lead to negatively charge O-planes, as they can satisfy the tadpole cancellation condition. Now one only has to compare the B-branes classes of O-planes with the point class to figure out which orientifold projection lead to O-planes with D3 charges, and compute the corresponding D3-tadpole charge. Importantly, note that the factor between the O-plane class and the point class has to be multiplied by four to obtain the tadpole charge of the O-planes to account for the extended spatial dimensions \cite{Becker_2007}.

When an ``N/A'' appears in the table, that means the O-plane and fluxes charges cannot cancel each other, since they belong to different topological classes. So there is no good geometric interpretation of the tadpole cancellation condition at this point in the moduli space, and one shall not consider these orientifold compactifications when fluxes are included as background.

\begin{table}[t]
  \centering
  \caption{Orientifolds in the $1^3 \times 4^3_A \times 0^3$ model}\label{1^3-4^3A-0^3}
    \begin{tabular}{|c|c|c|c|}
    \hline
    parity & $h^{2,1}$ modes & O-plane class & O-plane charge \\
    \hline
    $\sigma_0;~\mathbf{m}=(0,0,0,0,0,0,0,0,0)$ & 83 & $ 5[\L_0]/2 $ & -10 \\
    \hline
    $\sigma_0;~\mathbf{m}=(0,0,0,3,3,3,0,0,0)$ & 83 & $ 3([\L_1]+[\L_2])/2 $ & N/A \\
    \hline
    $\sigma_1;~\mathbf{m}=(0,0,0,0,0,0,0,0,0)$ & 62 & $ -5([\L_1] + [\L_2])/6 $ & N/A \\
    \hline
    $\sigma_1;~\mathbf{m}=(0,0,0,3,3,3,0,0,0)$ & 62 & $ -3[\L_0]/2 $ & -6 \\
    \hline
    $\sigma_2;~\mathbf{m}=(0,0,0,0,0,0,0,0,0)$ & 51 & $-([\L_1] + [\L_2])/2$ & N/A \\
    \hline
    $\sigma_2;~\mathbf{m}=(0,0,0,0,0,3,0,0,0)$ & 46 & $ [\L_0]/2 $ & -2 \\
    \hline
    $\sigma_3;~\mathbf{m}=(0,0,0,0,0,0,0,0,0)$ & 46 & $ [\L_0]/2 $ & -2 \\
    \hline
    $\sigma_3;~\mathbf{m}=(0,0,0,0,0,3,0,0,0)$ & 46 & $ -([\L_1] + [\L_2])/6 $ & N/A \\
    \hline
    $\Tilde{\sigma_0};~\mathbf{m}=(0,0,0,0,0,0,0,0,0)$ & 84 & $3([\L_0] + [\L_1])/2$ & N/A \\
    \hline
    $\Tilde{\sigma_0};~\mathbf{m}=(0,0,0,3,3,3,0,0,0)$ & 84 & $-5[\L_2]/2$ & N/A \\
    \hline
    $\Tilde{\sigma_1};~\mathbf{m}=(0,0,0,0,0,0,0,0,0)$ & 63 & $ -3[\L_2]/2 $ & N/A \\
    \hline
    $\Tilde{\sigma_1};~\mathbf{m}=(0,0,0,3,3,3,0,0,0)$ & 63 & $ -5([\L_0]+[\L_1])/6 $ & N/A \\
    \hline
    $\Tilde{\sigma_2};~\mathbf{m}=(0,0,0,0,0,0,0,0,0)$ & 52 & $ -[\L_2]/2 $ & N/A \\
    \hline
    $\Tilde{\sigma_2};~\mathbf{m}=(0,0,0,0,0,3,0,0,0)$ & 47 & $ -([\L_0] + [\L_1])/2 $ & N/A \\
    \hline
    $\Tilde{\sigma_3};~\mathbf{m}=(0,0,0,0,0,0,0,0,0)$ & 47 & $ -([\L_0] + [\L_1])/6 $ & N/A \\
    \hline
    $\Tilde{\sigma_3};~\mathbf{m}=(0,0,0,0,0,3,0,0,0)$ & 47 & $[\L_2]/2$ & N/A \\
    \hline
  \end{tabular}
\end{table}

The same thing can be done with the other geometric realization of this LG orbifold, with only one trivial minimal model factor. It is the mirror of a $T^2\times K3/\mathbb{Z}_3$ orbifold with point class
\begin{equation}
    [\mathrm{pt.}]_{T^2\times K3}=\frac{[\Lambda_1]+[\Lambda_2]}{3}=\frac{2[\Lambda_1]-[\Lambda_0]}{3}.
\end{equation}

So far, we do not have a clear criterion to single out a unique geometric interpretation. This ambiguity should not be viewed as a problem of selecting the ``correct'' geometry, but rather as a manifestation of the fact that the underlying SCFT admits multiple inequivalent geometric realizations. In particular, while the correspondence between different phases of the GLSM and the resulting IR conformal field theory is well-defined, the maps relating LG models and Calabi–Yau geometries are not one-to-one. Consequently, distinct geometries (sometimes with different topological structures) may correspond to the same conformal field theory, leading to what we refer to as ``ambiguous geometry''. This is closely analogous to the existence of multiple large volume limits in moduli space, or to different duality frames describing the same physical theory.\\

\begin{table}[t]
  \centering
  \caption{Orientifolds in the $1^3\times4^3_A\times0$ model}\label{1^3-4^3A-0}
    \begin{tabular}{|c|c|c|c|}
    \hline
    parity & $h^{2,1}$ modes & O-plane class & O-plane charge \\
    \hline
    $\sigma_0;~\mathbf{m}=(0,0,0,0,0,0,0)$ & 83 & $-3([\L_1] + [\L_2])$ & -36 \\
    \hline
    $\sigma_0;~\mathbf{m}=(0,0,0,3,3,3,0)$ & 83 & $5[\L_0]$ & N/A \\
    \hline
    $\sigma_1;~\mathbf{m}=(0,0,0,0,0,0,0)$ & 62 & $3[\L_0]$ & N/A \\
    \hline
    $\sigma_1;~\mathbf{m}=(0,0,0,3,3,3,0)$ & 62 & $-5([\L_1] + [\L_2])/3$ & -20 \\
    \hline
    $\sigma_2;~\mathbf{m}=(0,0,0,0,0,0,0)$ & 51 & $[\L_0]$ & -4 \\
    \hline
    $\sigma_2;~\mathbf{m}=(0,0,0,0,0,3,0)$ & 46 & $-[\L_1] - [\L_2]$ & -12 \\
    \hline
    $\sigma_3;~\mathbf{m}=(0,0,0,0,0,0,0)$ & 46 & $-([\L_1] + [\L_2])/3$ & -4 \\
    \hline
    $\sigma_3;~\mathbf{m}=(0,0,0,0,0,3,0)$ & 46 & $-[\L_0]$ & N/A \\
    \hline
    $\Tilde{\sigma_0};~\mathbf{m}=(0,0,0,0,0,0,0)$ & 84 & $5[\L_2]$ & N/A \\
    \hline
    $\Tilde{\sigma_0};~\mathbf{m}=(0,0,0,3,3,3,0)$ & 84 & $3([\L_0]+[\L_1])$ & N/A \\
    \hline
    $\Tilde{\sigma_1};~\mathbf{m}=(0,0,0,0,0,0,0)$ & 63 & $5([\L_0] + [\L_1])/3$ & N/A \\
    \hline
    $\Tilde{\sigma_1};~\mathbf{m}=(0,0,0,3,3,3,0)$ & 63 & $-3[\L_2]$ & N/A \\
    \hline
    $\Tilde{\sigma_2};~\mathbf{m}=(0,0,0,0,0,0,0)$ & 52 & $[\L_0] + [\L_1]$ & N/A \\
    \hline
    $\Tilde{\sigma_2};~\mathbf{m}=(0,0,0,0,0,3,0)$ & 47 & $[\L_2]$ & N/A \\
    \hline
    $\Tilde{\sigma_3};~\mathbf{m}=(0,0,0,0,0,0,0)$ & 47 & $-[\L_2]$ & N/A \\
    \hline
    $\Tilde{\sigma_3};~\mathbf{m}=(0,0,0,0,0,3,0)$ & 47 & $([\L_0] + [\L_1])/3$ & N/A \\
    \hline
  \end{tabular}
\end{table}

\noindent \textbf{Example: $1^3 \times 4^3_D$ model}\\

\noindent There now exists another symmetry for the $4_D$ minimal model, i.e., $(y,z)\rightarrow(y,-z)$, which we will denote by a minus sign in the corresponding entries of $\mathbf{m}$ (in the trivial case, we instead use a plus sign). It is also interesting to see that a switch between two $1_A$ variables is equivalent to $(y,z)\rightarrow(y,-z)$, which is essentially due to the isomorphism between $4_D$ and $1_A\times1_A$ SCFTs \cite{FUCHS1989317,Fuchs:1989yv}. An explicit way to see this is that by field redefinition, namely, setting
\begin{equation}
    \left\{ \begin{array}{ll} x_1=2^{-\frac{1}{3}}y+\frac{\sqrt{3}}{3}2^{-\frac{1}{3}}z \\  x_2=2^{-\frac{1}{3}}y-\frac{\sqrt{3}}{3}2^{-\frac{1}{3}}z \end{array} \right. ~ \Leftrightarrow ~ \left\{ \begin{array}{ll} y=2^{-\frac{2}{3}}(x_1+x_2) \\ z= 2^{-\frac{2}{3}} \sqrt{3} (x_1-x_2) \end{array} \right. .
\end{equation}
Note that this proves the equivalence $4_D \cong 1_A\times1_A$ in the closed string sector only. Considerations on D-branes and O-planes would require the equivalence to be true in the open string sector as well, but we have not been able to find strong motivations to support this conjecture so far.

One can also permute variables between two $4_D$ models, namely, simultaneously set $y_m\leftrightarrow y_n$, $z_m\leftrightarrow z_n$. Note that switching two $4_D$ models is equivalent to switching them while simultaneously putting a negative sign in front of both $z$'s up to a redefinition of variables and perhaps a proper change of superpotential. To be explicit, redefine $(\tilde{y_1},\tilde{z_1},\tilde{y_2},\tilde{z_2})=(y_1,\pm z_2,y_2,\pm z_1)$ to see that $\tilde{\mathcal{W}}=\tilde{y_1}^3+\tilde{y_1}\tilde{z_2}^2+\tilde{y_2}^3+\tilde{y_2}\tilde{z_1}^2$, the action $(y_1,z_1)\leftrightarrow(y_2,z_2)$ is therefore equivalent to $(\tilde{y_1},\tilde{z_1})\leftrightarrow(\tilde{y_2},-\tilde{z_2})$. For convenience, we order the LG variables as $\{x_1, x_2, x_3, y_1, y_2, y_3, z_1, z_2, z_3\}$ and denote permutations as $\sigma_0=\mathrm{id}$, $\sigma_1=(12)$, $\sigma_2=(45)(78)$ and $\sigma_3=(12)(45)(78)$. Note that the equation of motion for $y$ does not allow switch between $y$ and $z$ although they have the same weight. The O-plane classes are listed in Table \ref{O1^3-4^3D}.

\begin{table}[t]
  \centering
  \caption{Orientifolds in the $1^3\times4^3_D$ model}\label{O1^3-4^3D}
    \begin{tabular}{|c|c|c|c|}
    \hline
    parity & $h^{2,1}$ modes & O-plane class \\
    \hline
    $\sigma_0;~\mathbf{m}=(0,...,0,+,+,+)$ & 84 & $3([\L_1] - [\L_2])$ \\
    \hline
    $\sigma_1;~\mathbf{m}=(0,...,0,+,+,+)$ & 63 & $3[\L_0]$ \\
    \hline
    $\sigma_2;~\mathbf{m}=(0,...,0,+,+,+)$ & 52 & $\lt - \lo$ \\
    \hline
    $\sigma_3;~\mathbf{m}=(0,...,0,+,+,+)$ & 47 & $-\lz$ \\
    \hline
    $\sigma_3;~\mathbf{m}=(0,...,0,+,+,-)$ & 44 & $(\lo - \lt)/3$ \\
    \hline
    $\Tilde{\sigma_0};~\mathbf{m}=(0,...,0,+,+,+)$ & 84 & $3([\L_1] - [\L_0])$ \\
    \hline
    $\Tilde{\sigma_1};~\mathbf{m}=(0,...,0,+,+,+)$ & 63 & $-3[\L_2]$ \\
    \hline
    $\Tilde{\sigma_2};~\mathbf{m}=(0,...,0,+,+,+)$ & 52 & $\lz - \lo$ \\
    \hline
    $\Tilde{\sigma_3};~\mathbf{m}=(0,...,0,+,+,+)$ & 47 & $\lt$ \\
    \hline
    $\Tilde{\sigma_3};~\mathbf{m}=(0,...,0,+,+,-)$ & 44 & $(\lo - \lz)/3$ \\
    \hline
  \end{tabular}
\end{table}

There are again two ways to understand the Calabi-Yau geometry in the large volume. When considered as the tensor product of three identical $1\times4_D$ models (and with a $\mathbb{Z}_3\times\mathbb{Z}_3$ projection), the natural matrix factorization of the $1\times4_D$ model (with an LG potential $\mathcal{W}=x^3+y^3+yz^2$) is
\begin{equation}
    Q=\begin{pmatrix}
0 & x \\
x^2 & 0
\end{pmatrix}\otimes\mathbb{1}+\sigma\otimes\
\begin{pmatrix}
0 & y \\
y^2+z^2 & 0 
\end{pmatrix}.
\end{equation}
This happens to give a minimal basis but does not give a D0-brane. However the polynomial $x^3+y^3+yz^2$ cannot be factorized, so we do not have a matrix factorization description of a D0-brane using permutation branes for this internal space. 

The other way to construct the geometry is to consider the model as a $\mathbb{Z}_3$ orbifolded tensor product of a $1^3$ model and a $4^3_D$ model. Let us focus on the $4^3_D$ geometry. The LG superpotential is
 \begin{equation}
     \mathcal{W} = y_1^3 + y_1 z_1^2 + y_2^3 + y_2 z_2^2 + y_1^3 + y_3 z_3^2.
 \end{equation}
 The corresponding Calabi-Yau is a (complex) codimension 2 surface in $\mathbb{CP}^3\times\mathbb{CP}^1$ defined by
 \begin{equation}
\left\{
\begin{array}{l}
y_1^3+y_2^3+y_3^3+y_1z_1^2=0 \\
y_2z_2^2+y_3z_3^2=0,
\end{array}
\right.
 \end{equation}
where $[y_1:y_2:y_3:z_1]$ and $[z_2:z_3]$ are homogeneous coordinates of $\mathbb{CP}^3$ and $\mathbb{CP}^1$, respectively. Just like with the $T^6$ interpretation of this Gepner model, it was impossible to find a matrix factorization contruction of a basis of branes containing a pure D0-brane state.\\

\noindent \textbf{Example: $2^6$ model}\\

\begin{table}[t]
  \centering
  \caption{Orientifolds in the $2^6\times0$ model}\label{2^6-0}
    \begin{tabular}{|c|c|c|c|}
    \hline
    parity & $h^{2,1}$ modes & O-plane class & O-plane charge \\
    \hline
    $\sigma_0$ & 90 & $-7[\Lambda_1]/2$ & -28\\
    \hline
    $\sigma_1$ & 60 & $3[\Lambda_0]/2$ & N/A\\
    \hline
    $\sigma_2$ & 50 & $-[\Lambda_1]/2$ & -4\\
    \hline
    $\sigma_3$ & 48 & $-[\Lambda_0]/2$ & N/A\\
    \hline
    $\Tilde{\sigma_0}$ & 90 & $5(-[\Lambda_0]+[\Lambda_1])/2$ & N/A\\
    \hline
    $\Tilde{\sigma_1}$ & 60 & $[\Lambda_0]+[\Lambda_1]$ & N/A\\
    \hline
    $\Tilde{\sigma_2}$ & 50 & $([\Lambda_0]-[\Lambda_1])/2$ & N/A\\
    \hline
    $\Tilde{\sigma_3}$ & 48 & 0 & 0\\
    \hline
  \end{tabular}
\end{table}

\noindent We mention this last example because we found different results from \cite{Becker_2007}. The non-equivalent involutions of this model, with superpotential $\W = x_1^4 + ... + x_6^4 + z_1^2 ~ (+ z_2^2 + z_3^2)$, are $\sigma_1=(12)$, $\sigma_2=(12)(34)$ and $\sigma_3=(12)(34)(56)$. Again, for $2^6$ models with three trivial factors, there is freedom to permute two of the trivial variables, which turns out to be a relabeling of ordinary branes, and therefore unnecessary to discuss separately below. They are dressed by trivial or non-trivial quantum symmetry action, and there is no non-trivial phase $\mathbf{m}$. There are two possible interpretations again. The first one is $2^6\times0$ leading to the mirror of $T^2\times K3/\Z_4$. The B-branes in this model have charges
\begin{equation}
    \braket{\L_n}{k} = 2 i^{kn} (1-i^k)^6,
\end{equation}
while crosscap states (with the involution $\sigma_0$) and potentially corresponding O-planes have charges and classes ($k=1,3$)
\begin{equation}
    \begin{aligned}
        \braket{C}{k} & = -56 & ~ ; ~ [O_0] & = \frac{7 [\L_1]}{2} \\
        \langle \tilde{C} | k \rangle & = -40(1+i^k) & ~ ; ~ [\tilde{O}_0] & = \frac{5([\L_1] - [\L_0])}{2}
    \end{aligned}
\end{equation}
where $|\tilde{C} \rangle$ is the crosscap state of the involution dressed by quantum symmetry. The $T^2$ and $K3$ have respective D0-branes charges (for $k=1,3$) 
\begin{equation}
        \langle [pt.]_{T^2} | k \rangle = 2(1-i^k) ~ \text{ and }  ~ \langle [pt.]_{K3} | k \rangle = (1-i^k)^3
\end{equation}
so that the orbifolded $T^2\times K3/\Z_4$ has point class
\begin{equation}
    \begin{aligned}
        \langle [pt.]_{T^2\times K3/\Z_4} | k \rangle & = 2(1-i^k)^4~, \\
        [pt.]_{T^2\times K3/\Z_4} & = \pm \frac{[\L_1]}{2}~.
    \end{aligned}
\end{equation}
It turns out that the D3-brane tadpole cancellation condition in these orientifold models only makes sense in the case of involutions without dressing by quantum symmetry. The resulting O-planes have a spacetime D3-brane class
\begin{equation}
    [O_0] = 7 [pt.]_{T^2\times K3/\Z_4}~.
\end{equation}
When mutliplied by 4 to account for the extended spacetime directions, we obtain a D3-tadpole charge of $-28$ for this O-plane.

The second geometric interpretation is the $2^6\times0^3$ model, which corresponds to $T^6/\Z_4\times \Z_4$. The additionnal two quadratic fields multiply the charges of the B-branes $\L_n$ by 4, whereas they only multiply the crosscap state charges by $2i$ (this is one aspect of the phenomenon of extended Knörrer periodicity, see \cite{Walcher_2005, Brunner_2004})
\begin{equation}
    [\hat{\L}_n] = 4 [\L_n] ~ \text{ and } ~ [\hat{O}_0] = 2i[O_0]~.
\end{equation}
The three $T^2$ that constitute the $T^6 = T^2 \times T^2 \times T^2$ are the same ones as in the $2^6 \times 0$ interpretation, so that we can easily compute the D0-brane charge and class of the orbifolded $T^6/\Z_4\times \Z_4$ to be
\begin{equation}
    \begin{aligned}
        \langle [pt.]_{T^6/\Z_4 \times \Z_4} | k \rangle & = 8(1-i^k)^3 \\
        [pt.]_{T^6/\Z_4 \times \Z_4} & = \frac{-[\hat{\L}_0] \pm [\hat{\L}_1]}{4}~.
    \end{aligned}
\end{equation}
This time, only the orientifold involution dressed by quantum symmetry has charges compatible with the D0-branes class, namely
\begin{equation}
    [\hat{\tilde{O}}_0] = 5 [pt.]_{T^6/\Z_4 \times \Z_4}~.
\end{equation}
From this, we find the D3 tadpole charge of the O-plane corresponding to the trivial permutation to be 20, compared to the 40 derived in \cite{Becker_2007}.

The O-planes classes and tadpole charges for the $2^6\times0$ and $2^6\times0^3$ models are given in \autoref{2^6-0} and \autoref{2^6-0^3} respectively. One notable consequence of the difference of a factor of 2 between our results and those of \cite{Becker_2007} for the tadpole bound of the $2^6 \times 0^3$ model would be that the fully stabilized solutions found in \cite{Becker_2024_higherorder} and \cite{Becker:2024ayh} would now have a flux number above the tadpole bound, and are not physical solutions anymore. At this point, it seems that our best candidate for moduli stabilization is the orientifold with permutation $\sigma_0$ of the $1^3 \times 4^3 \times 0$ model, as it has the highest tadpole bound of the models without K{\"a}hler moduli that we studied.

\begin{table}[t]
  \centering
  \caption{Orientifolds in the $2^6\times0^3$ model}\label{2^6-0^3}
    \begin{tabular}{|c|c|c|c|}
    \hline
    parity & $h^{2,1}$ modes & O-plane class & O-plane charge \\
    \hline
    $\sigma_0$ & 90 & $-7[\Lambda_0]/4$ & N/A\\
    \hline
    $\sigma_1$ & 60 & $3[\Lambda_1]/4$ & N/A\\
    \hline
    $\sigma_2$ & 50 & $[\Lambda_0]/4$ & N/A\\
    \hline
    $\sigma_3$ & 48 & $-[\Lambda_1]/4$ & N/A\\
    \hline
    $\Tilde{\sigma_0}$ & 90 & $5([\Lambda_0]+[\Lambda_1])/4$ & -20\\
    \hline
    $\Tilde{\sigma_1}$ & 60 & $([\Lambda_0]-[\Lambda_1])/2$ & -8\\
    \hline
    $\Tilde{\sigma_2}$ & 50 & $([\Lambda_0]+[\Lambda_1])/4$ & -4\\
    \hline
    $\Tilde{\sigma_3}$ & 48 & 0 & 0\\
    \hline
  \end{tabular}
\end{table}

\section{Flux compactification, moduli stabilization and symmetries}
\label{sec:fluxquant}
As has been shown in \autoref{sec:review_Gepner}, three-form fluxes can either appear in untwisted or twisted sectors, and similarly for the complex structure moduli. Stabilization of twisted moduli with fluxes from twisted sectors has not been well understood so far, so for simplicity, we will only turn on fluxes from untwisted sectors and discuss flux quantization and tadpole cancellation conditions related to these fluxes. In the next section, we will discuss moduli stabilization with only ``untwisted fluxes'' turned on. For some discussion regarding the twisted fluxes and/or moduli, see \cite{Ishiguro:2024coq,Blumenhagen_2003,Blumenhagen_2005,antoniadis2024fluxvacuatypeiib}.

\subsection{Flux quantization}
Following the conventions in \cite{becker2023fluxes19landauginzburgmodel,Becker:2024ayh}, we give below the flux quantization condition formulated in the covering space $M$ (here covering space means the internal space before the orientifold projection),
\begin{equation}
\int_{\gamma_n}H_{\mathrm{NSNS}}\in\mathbb{Z},~~~~~~\int_{\gamma_n}H_{\mathrm{RR}}\in\mathbb{Z},
\end{equation}
where $\gamma_n\in H_3(M,\mathbb{Z})$. In other words, we work with a normalization for the fluxes so that the periods are integers. Note that there is no need to require the flux quanta in the covering space to be even, because O-planes can carry discrete fluxes so that the flux quantization in the orientifold space is also satisfied \cite{Frey_2002} (see \cite{Becker_2007} for some further discussions from the worldsheet point of view). Note that NSNS and RR fluxes transform as a doublet under the $SL(2,\mathbb{Z})$ modular group, which together with the invariance of the worldsheet superpotential $\mathcal{W}$ under the discrete symmetry considered in the theory, determines how the complex structure moduli and flux quanta transform, thus eventually impose a set of selection rules to the spacetime superpotential and even the K\"ahler potential. We will discuss this in more details in \autoref{sec:sta_sym}.

\subsection{Tadpole cancellation in the covering space}
In type IIB compactification, tadpole cancellation condition is the integrated form of the Bianchi identity for the 5-form field strength in the RR-sector \cite{Kachru_2003}. In the covering space, it reads \cite{Becker_2007},
\begin{equation}
    \int_MH_{\mathrm{RR}}\wedge H_{\mathrm{NSNS}}+N_{\mathrm{D3}}+Q_\mathrm{O3}=0,
    \label{eq:tadcond}
\end{equation}
where $N_{\mathrm{D3}}$ and $Q_\mathrm{O3}$ denote the number of spacetime filling D3-branes and the (negative) O3-plane charges, respectively. To be more precise, in the covering space, a D3-brane and its orientifold image are counted separately. A single ordinary (namely, non-exotic and carrying no discrete fluxes \cite{Frey_2002, Kachru_2003}) O3-plane contributes a charge of $-\frac{1}{2}$ instead of $-\frac{1}{4}$ as in the orientifold space. Assuming there are no background (anti-)$D3$-branes (thus no moduli introduced by them as well), the tadpole cancellation condition becomes
\begin{equation}
    N_{\mathrm{flux}}+Q_\mathrm{O3}=0,\label{tadpole cancellation}
\end{equation}
where $N_{\mathrm{flux}}=\int_MH_{\mathrm{RR}}\wedge H_{\mathrm{NSNS}}$. The LHS of (\ref{tadpole cancellation}) can be rewritten as
\begin{equation}
    N_{\mathrm{flux}}=\frac{1}{\tau-\bar{\tau}}\int_MG\wedge\bar{G_3},
\end{equation}
where $\tau$ is the axion-dilaton as before and $G_3=H_{\mathrm{RR}}-\tau H_{\mathrm{NSNS}}$ is the combined flux. After properly choosing a basis of 3-cycles and their Poincar\'e dual, one may use Riemann bilinear identity to further rewrite the above equation as a summation over quadratic terms of flux coefficients, as was done in \cite{Becker:2024ayh}.

\subsection{Moduli stabilization and symmetries}

\label{sec:sta_sym}

Since there are no K\"ahler moduli in the models discussed in this paper, moduli stabilization only concerns complex structure moduli, some of which are possibly from the twisted sectors. Turning off fluxes from twisted sectors will greatly simplify the moduli stabilization problem, because the twisted moduli will automatically become localized and stable \cite{antoniadis2024fluxvacuatypeiib,Cascales:2003zp}, at least at the quadratic level of correction. 

One may worry about the $\alpha'$ and the string coupling $g_s$ corrections of the flux induced superpotential in stabilizing the moduli, however, it was claimed in \cite{Becker_2007} that the superpotential does not receive any perturbative or non-perturbative corrections due to the fact that the LG models include all $\alpha'$ corrections and that the string coupling corrections are forbidden by a non-renormalization theorem (reasonably extended to the non-geometric cases). It should be noted that discrete symmetries can impose heavy constraints on the string vacua, as compared with points away from the symmetry loci. A recent study \cite{chen2025symmetriesmtheorylikevacuadimensions} purely based on the symmetry of the model, without assuming the validity of the non-renormalization theorem in the non-geometric case, claims that all the vacua survive all-order corrections to the superpotential, simply guaranteed by the symmetry selection rules for the expansion of $\mathcal{W}$. Such selection rules are established upon two facts: the GVW flux superpotential transforms with weight $-1$ under modular transformation and only the flux-dependent terms contribute to it linearly. The possible (moduli dependent) instanton corrections must also vanish because their modular transformation is incompatible with the complex structure moduli, which are neutral under modular group action. This implies that it is not necessary to assume any non-renormalization properties of the spacetime superpotential in the very beginning to find supersymmetric string vacua. In particular, when the axion-dilaton VEV is at the $\mathbb{Z}_4$ ($\tau=i$) or $\mathbb{Z}_3$ ($\tau=\mathrm{e}^\frac{2\pi i}{3}$) symmetry point, and requiring that there are no new
additional massless fields at these points, it has been shown that the sign of vacuum energy can be determined simply by the weight of superpotential under $SL(2,\mathbb{Z})$ modular transformation, and that criticality is automatically guaranteed \cite{Mohseni:2025tig}.

Nonetheless, one should expect that in the geometric LG models (or even some non-geometric ones), the symmetry may be reduced to a much smaller subgroup and will not be powerful enough to impose strong constraints, especially when some particular orientifolds are considered. An example of this in the geometric case is that type-B orientifold projection selects a real subspace of K\"ahler moduli space and the discrete symmetry group can be no larger than some power of $\mathbb{Z}_2$'s. More mathematical viewpoints and arguments about discrete symmetries in modular space are given for example in \cite{Grimm:2024fip}, where the authors showed that along certain symmetry loci in moduli space, the generically transcendental scalar potentials become algebraic and lead to a finite exact vacua after imposing the tadpole bound. It should be interesting to explicitly verify some of these properties due to discrete symmetry by solving any non-geometric models discussed in this paper.

\section{Summary and outlook}
In this work, we have classified B-type orientifolds of Gepner models without Kähler moduli and derived the corresponding O-plane charges and tadpole cancellation conditions. These results provide a foundation for further studies of flux compactifications in non-geometric backgrounds and their associated moduli stabilization mechanisms.

Several open directions remain to be explored. One important question concerns the precise characterization of D-brane charge lattices in Gepner models involving D-type minimal-model factors. While permutation branes provide a convenient minimal basis in many A-type models and allow for the identification of D0-brane charges, the situation is considerably more subtle in the D-type case. In particular, there exist D0-brane charge classes that cannot be obtained from the standard permutation-brane construction. It would therefore be interesting to develop a systematic matrix-factorization description of these branes and understand their role in the full charge lattice, especially in relation to flux-induced tadpoles and large-volume geometric interpretations.

Another natural direction is the study of twisted-sector moduli and fluxes in models with D-type modular invariants. Since a significant fraction of the non-geometric Gepner models with $h^{1,1}=0$ belong to this class, a better understanding of their twisted sectors may lead to new classes of stabilized vacua. It would also be worthwhile to investigate whether the enhanced discrete symmetries present in many of these models can be used to constrain non-perturbative effects or provide alternative explanations for moduli stabilization.

More broadly, the interplay between orientifolds, fluxes, matrix factorizations, and LG-CY correspondence in non-geometric compactifications remains only partially understood. We hope that the results presented here will serve as a useful starting point for a more complete treatment of D-branes and flux vacua in Gepner-model compactifications beyond the geometric regime.

\section*{Acknowledgements}

We would like to thank Antoine Bourget, Dan Israël, Quentin Lamouret, Muthusamy Rajaguru, Anindya Sengupta, Sav Sethi, Eric Sharpe and Thibaud Raymond for useful discussions. We are especially thankful to Katrin Becker and Mariana Graña for their essential help in the realization of this paper. The work of MM is supported by the {\'E}cole Normale Sup{\'e}rieure de Lyon via a CDSN doctoral grant. The work of JW is funded by the Deutsche Forschungsgemeinschaft (DFG, German Research Foundation) under Germany’s Excellence Strategy EXC 2181/1 — 390900948 (the Heidelberg STRUCTURES Excellence Cluster). The work of QY is supported by the NSF grant PHY-2413006 and endowment funds from the Mitchell Family Foundation.

\newpage

\appendix

\section*{Appendix}

\section{(2,1)-deformations of the LG models with \texorpdfstring{$h^{1,1}=0$}{h11=0}}

\label{app:RRgs}

We give the Hodge numbers and complex structure deformations in terms of LG variables of the models that have not been treated in the core of the text.\\

\begin{minipage}{0.5\textwidth}
    \centering
    \captionof{table}{$1^7\times4_A$ model, $h^{2,1} = 84$ \vspace{0.3cm} \\ \centerline{$\mathcal{W} = \sum_{i=1}^7 x_i^3 + y^6$}} \label{1^7-4A}
    \begin{tabular}{|c|c|c|}
        \hline
        sector & monomials & \# \\
        \hline
        \multirow{3}{*}{untwisted} & $x_ix_jx_k$ & 35 \\
                            & $x_ix_jy^2$ & 21 \\
                            & $x_iy^4$ & 7 \\
        \hline
        $\nu=3$ twisted & $x_i x_j$ & 21 \\
        \hline
    \end{tabular}

    \vspace{1cm}

    \centering
    \captionof{table}{$1^5\times4_A\times4_D$ model, $h^{2,1}=84$ \vspace{0.3cm} \\ \centerline{$\mathcal{W} = \sum_{i=1}^5 x_i^3 + y^6 + p^3 + pq^2$}} \vspace{0.3cm} \label{1^5-4A-4D}
    \begin{tabular}{|c|c|c|}
        \hline
        sector & monomials & \# \\
        \hline
        \multirow{11}{*}{untwisted} & $x_ix_jx_k$ & 10 \\
                            & $x_ix_jy^2$ & 10 \\
                            & $x_ix_jp$ & 10 \\
                            & $x_ix_jq$ & 10 \\
                            & $x_iy^4$ & 5 \\
                            & $x_ip^2$ & 5 \\
                            & $x_iy^2p$ & 5 \\
                            & $x_iy^2q$ & 5 \\
                            & $y^4p$ & 1 \\
                            & $y^4q$ & 1 \\
                            & $y^2p^2$ & 1 \\
        \hline
        \multirow{4}{*}{$\nu=3$ twisted} & $x_i x_j$ & 10 \\
        & $x_i p$ & 5 \\
        & $x_i q$ & 5 \\
        & $p^2$ & 1 \\
        \hline
    \end{tabular}
\end{minipage}
\begin{minipage}{0.5\textwidth}
    \centering
    \captionof{table}{$1^7\times4_D$ model, $h^{2,1} = 84$ \vspace{0.3cm} \\ \centerline{$\mathcal{W} = \sum_{i=1}^7 x_i^3 + y^3 + y z^2$}} \label{1^7-4D}
    \begin{tabular}{|c|c|c|}
        \hline
        sector & monomials & \# \\
        \hline
        \multirow{4}{*}{untwisted} & $x_ix_jx_k$ & 35 \\
                            & $x_ix_jy$ & 21 \\
                            & $x_ix_jz$ & 21 \\
                            & $x_iy^2$ & 7 \\
        \hline
    \end{tabular}

    \vspace{1cm}

    \centering
    \captionof{table}{$1^5\times4_D^2$ model, $h^{2,1} = 84$ \vspace{0.3cm} \\ \centerline{$\mathcal{W} = \sum_{i=1}^5 x_i^3 + \sum_{m=1}^2 (y_m^3 + y_m z_m^2)$}} \vspace{0.3cm} \label{1^5-4^2D}
    \begin{tabular}{|c|c|c|}
        \hline
        sector & monomials & \# \\
        \hline
        \multirow{9}{*}{untwisted} & $x_ix_jx_k$ & 10 \\
                            & $x_ix_jy_m$ & 20 \\
                            & $x_ix_jz_m$ & 20 \\
                            & $x_iy_m^2$ & 10 \\
                            & $x_iy_mz_n$ & 10 \\
                            & $x_iy_1y_2$ & 5 \\
                            & $x_iz_1z_2$ & 5 \\
                            & $y_m^2y_n$ & 2 \\
                            & $y_m^2z_n$ & 2 \\
        \hline
    \end{tabular}
\end{minipage}

\begin{minipage}{0.5\textwidth}
    \centering
    \captionof{table}{$1^3\times4_A\times4^2_D$ model, $h^{2,1} = 84$ \vspace{0.3cm} \\ \centerline{$\mathcal{W}=\sum_{i=1}^3x_i^3+y^6+\sum_{m=1}^2(p_m^3+p_mq_m^2)$}}\label{1^3-4A-4^2D}
    \begin{tabular}{|c|c|c|}
        \hline
        sector & monomials & \# \\
        \hline
        \multirow{17}{*}{untwisted} & $x_1x_2x_3$ & 1 \\
                            & $x_ix_jy^2$ & 3 \\
                            & $x_ix_jp_m$ & 6 \\
                            & $x_ix_jq_m$ & 6 \\
                            & $x_iy^4$ & 3 \\
                            & $x_iy^2p_m$ & 6 \\
                            & $x_iy^2q_m$ & 6 \\
                            & $x_ip_m^2$ & 6 \\
                            & $x_ip_1p_2$ & 3 \\
                            & $x_iq_1q_2$ & 3 \\
                            & $x_ip_mq_n$ & 6 \\
                            & $y^4p_m$ & 2 \\
                            & $y^4q_m$ & 2 \\
                            & $y^2p_m^2$ & 2 \\
                            & $y^2p_1p_2$ & 1 \\
                            & $y^2q_1q_2$ & 1 \\
                            & $y^2p_mq_n$ & 2 \\
                            & $p_m^2p_n$ & 2 \\
                            & $p_m^2q_n$ & 2 \\
        \hline
        \multirow{7}{*}{$\nu=3$ twisted} & $x_i x_j$ & 3 \\
         & $x_ip_m$ & 6 \\
        & $x_iq_m$ & 6 \\
        & $p_m^2$ & 2 \\
        & $p_1p_2$ & 1 \\
        & $q_1q_2$ & 1 \\
        & $p_mq_n$ & 2 \\
        \hline
    \end{tabular}
    
    \vspace{1cm}

    \centering
    \captionof{table}{$1^9$ model, $h^{2,1}=84$\vspace{0.3cm} \\ \centerline{$\mathcal{W} = \sum_{i=1}^9 x_i^3$}} \vspace{0.3cm} \label{1^9}
    \begin{tabular}{|c|c|c|}
        \hline
        sector & monomials & \# \\
        \hline
        untwisted & $x_i x_j x_k$ & 84 \\
        \hline
    \end{tabular}  
\end{minipage}
\begin{minipage}{0.5\textwidth}
    \centering
    \captionof{table}{$1\times4^3_A\times4_D$ model, $h^{2.1}=84$ \vspace{0.3cm} \\ \centerline{$\mathcal{W} = x^3 + \sum_{i=1}^3 y_i^6 + p^3 + p q^2$}} \label{1-4^3A-4D}
    \begin{tabular}{|c|c|c|}
        \hline
        sector & monomials & \# \\
        \hline
        \multirow{19}{*}{untwisted} & $xy_i^4$ & 3 \\
                            & $xy_i^3y_j$ & 6 \\
                            & $xy_i^2y_j^2$ & 3 \\
                            & $xy_i^2y_jy_k$ & 3 \\
                            & $xy_i^2p$ & 3 \\
                            & $xy_i^2q$ & 3 \\
                             & $xy_iy_jp$ & 3 \\
                            & $xy_iy_jq$ & 3 \\
                            & $xp^2$ & 1 \\
                            & $y_i^4y_j^2$ & 6 \\
                            & $y_i^4y_jy_k$ & 3 \\
                            & $y_i^3y_j^3$ & 3 \\
                            & $y_i^3y_j^2y_k$ & 6 \\
                            & $y_1^2y_2^2y_3^2$ & 1 \\
                            & $y_i^4p$ & 3 \\
                            & $y_i^4q$ & 3 \\
                            & $y_i^3y_jp$ & 6 \\
                            & $y_i^3y_jq$ & 6 \\
                            & $y_i^2y_j^2p$ & 3 \\
                            & $y_i^2y_j^2q$ & 3 \\
                            & $y_i^2y_jy_kp$ & 3 \\
                            & $y_i^2y_jy_kq$ & 3 \\
                            & $y_i^2p^2$ & 3 \\
                            & $y_iy_jp^2$ & 3 \\
        \hline
        $\nu=3$ twisted & 1 & 1 \\                 
        \hline
    \end{tabular}
      
    \vspace{1cm}

    \centering
    \captionof{table}{$2^6$ model, $h^{2,1}=90$\vspace{0.3cm} \\ \centerline{$\mathcal{W} = \sum_{i=1}^6 x_i^4$}} \vspace{0.3cm} \label{2^6}
    \begin{tabular}{|c|c|c|}
        \hline
        sector & monomials & \# \\
        \hline
        \multirow{3}{*}{untwisted} & $x_i x_j x_k x_l$ & 15 \\
        & $x_i x_j x_k^2$ & 60 \\
        & $x_i^2 x_j^2$ & 15 \\
        \hline
    \end{tabular} 
\end{minipage}

\begin{minipage}{0.5\textwidth}
    \centering
    \captionof{table}{$1\times4_A\times4^3_D$ model, $h^{2,1}=84$ \vspace{0.3cm} \\ \centerline{$\mathcal{W}=x^3+y^6+\sum_{m=1}^3(p_m^3+p_mq_m^2)$}}\label{1-4A-4^3D}
    \begin{tabular}{|c|c|c|}
        \hline
        sector & monomials & \# \\
        \hline
        \multirow{19}{*}{untwisted} & $xy^4$ & 1 \\
                            & $xy^2p_m$ & 3 \\
                            & $xy^2q_m$ & 3 \\
                            & $xp_mq_n$ & 6 \\
                            & $xp_m^2$ & 3 \\
                            & $xp_mp_n$ & 3 \\
                            & $xq_mq_n$ & 3 \\
                            & $y^4p_m$ & 3 \\
                            & $y^4q_m$ & 3 \\
                            & $y^2p_m^2$ & 3 \\
                            & $y^2p_mp_n$ & 3 \\
                            & $y^2p_mq_n$ & 6 \\
                            & $y^2q_mq_n$ & 3 \\
                            & $p_m^2p_n$ & 6 \\
                            & $p_m^2q_n$ & 6 \\
                            & $p_mp_nq_l$ & 3 \\
                            & $p_mq_nq_l$ & 3 \\
                            & $p_1p_2p_3$ & 1 \\
                            & $q_1q_2q_3$ & 1 \\
        \hline
        \multirow{6}{*}{$\nu=3$ twisted} & $x p_m$ & 3 \\
        & $xq_m$ & 3 \\
        & $p_m^2$ & 3 \\
        & $p_mp_n$ & 3 \\
        & $p_mq_n$ & 6 \\
        & $q_mq_n$ & 3 \\
        \hline
    \end{tabular}
\end{minipage}
\begin{minipage}{0.5\textwidth}
    \centering
    \captionof{table}{$1\times4^4_D$ model, $h^{2,1}=84$ \vspace{0.3cm} \\ \centerline{$\mathcal{W} = x^3 + \sum_{i=1}^4 (y_i^3 + y_i z_i^2)$}}
    \label{1-4^4D}
    \begin{tabular}{|c|c|c|}
        \hline
        sector & monomials & \# \\
        \hline
        \multirow{10}{*}{untwisted} & $xp_i^2$ & 4 \\
                            & $xp_ip_j$ & 6 \\
                            & $xp_iq_j$ & 12 \\
                            & $xq_iq_j$ & 6 \\
                            & $p_i^2p_j$ & 12 \\
                            & $p_i^2q_j$ & 12 \\
                            & $p_ip_jp_k$ & 4 \\
                            & $p_ip_jq_k$ & 12 \\
                            & $p_iq_jq_k$ & 12 \\
                            & $q_iq_jq_k$ & 4 \\                 
        \hline
    \end{tabular}
\end{minipage}

\section{O-planes charges in the LG models with \texorpdfstring{$h^{1,1}=0$}{h11=0}}

\label{app:Oplanes}

We give the orientifold projections, B-branes classes and when available the tadpole charges of the associated O-planes of the models that have not been treated in the core of the text. $\sigma_0$ always designates the trivial permutation and $\tilde{\sigma}_i$ designates the involution dressed by quantum symmetry. The phase happens to be $e^{i\pi/3}$ for all of the following models. We also include the phases $\mathbf{m}$ that are not equivalent to an orbifold action.\\

\newpage

\noindent $1^9$ model with variables $(x_1,...,x_9)$
\begin{itemize}
    \item $\sigma_1=(1 2) ~ ; ~ \sigma_2=(1 2)(3 4) ~ ; ~ \sigma_3=(1 2)(3 4)(5 6) ~ ; ~ \sigma_4=(1 2)(3 4)(5 6)(7 8)$
    \item $(T^2 : \{x_i^3 + x_j^3 + x_k^3 = 0\} \subset \CP^2)^3/\Z_3 \times \Z_3$ \footnote{This $1^3$ torus will be simply noted $T^2$ in the following}
    \item $[\mathrm{pt.}]=\frac{[\Lambda_1]-[\Lambda_2]}{9}$
\end{itemize}
\begin{center}
\begin{tabular}{|c|c|c|c|}
    \hline
    parity & $h^{2,1}$ modes & O-plane class & O-plane charge \\
    \hline
    $\sigma_0$ & 84 & -$[\Lambda_0]$ & N/A\\
    \hline
    $\sigma_1$ & 63 & $(-[\Lambda_1]+[\Lambda_2])/3$ & -12\\
    \hline
    $\sigma_2$ & 52 & $[\Lambda_0]/3$ & N/A\\
    \hline
    $\sigma_3$ & 47 & $-([\Lambda_1]-[\Lambda_2])/9$ & -4\\
    \hline
    $\sigma_4$ & 44 & $-[\Lambda_0]/9$ & N/A\\  
    \hline
\end{tabular}
\end{center}
\vspace{0.5cm}

\noindent $1^7 \times 4_A$ model with variables $(x_1,...,x_7,y,z)$
\begin{itemize}
    \item $\sigma_1=(12) ~ ; ~ \sigma_2=(12)(34) ~ ; ~ \sigma_3=(12)(34)(56)$
    \item $(T^2)^2 \times (T^2 : \{x_7^3 + y^6 + z^2 = 0\} \subset \WCP^2[2,1,3])/\Z_6$
    \item $[\mathrm{pt.}]=\frac{[\Lambda_2]}{3}=\frac{[\Lambda_1]-[\Lambda_0]}{3}$
\end{itemize}
\begin{center}
\begin{tabular}{|c|c|c|c|}
    \hline
    parity & $h^{2,1}$ modes & O-plane class & O-plane charge \\
    \hline
    $\sigma_0;~\mathbf{m}=(0,...,0,0)$ & 63 & $-[\L_0]$ & -12 \\
    \hline
    $\sigma_1;~\mathbf{m}=(0,...,0,0)$ & 52 & $([\L_1] + [\L_2])/3$ & N/A \\
    \hline
    $\sigma_2;~\mathbf{m}=(0,...,0,0)$ & 47 & $-[\L_0]/3$ & -4 \\ 
    \hline
    $\sigma_3;~\mathbf{m}=(0,...,0,0)$ & 44 & $-(\lo+\lt)/9$ & N/A \\
    \hline
    $\sigma_0;~\mathbf{m}=(0,...,3,0)$ & 63 & $[\L_1]+[\L_2]$ & N/A \\
    \hline
    $\sigma_1;~\mathbf{m}=(0,...,3,0)$ & 52 & $-[\L_0]$ & -12 \\
    \hline
    $\sigma_2;~\mathbf{m}=(0,...,3,0)$ & 47 & $([\L_1] + [\L_2])/3$ & N/A \\ 
    \hline
    $\sigma_3;~\mathbf{m}=(0,...,3,0)$ & 44 & $-[\L_0]/3$ & -4 \\
    \hline
    $\Tilde{\sigma_0};~\mathbf{m}=(0,...,0,0)$ & 84 & $-\lo-\lz$ & N/A \\
    \hline
    $\Tilde{\sigma_1};~\mathbf{m}=(0,...,0,0)$ & 63 & $-[\L_2]$ & -12 \\
    \hline
    $\Tilde{\sigma_2};~\mathbf{m}=(0,...,0,0)$ & 52 & $(\lz + \lo)/3$ & N/A \\
    \hline
    $\Tilde{\sigma_3};~\mathbf{m}=(0,...,0,0)$ & 47 & $-[\L_2]/3$ & -4 \\
    \hline
    $\Tilde{\sigma_0};~\mathbf{m}=(0,...,3,0)$ & 84 & $-\lt$ & -12 \\
    \hline
    $\Tilde{\sigma_1};~\mathbf{m}=(0,...,3,0)$ & 63 & $(\lz + \lo)/3$ & N/A \\
    \hline
    $\Tilde{\sigma_2};~\mathbf{m}=(0,...,3,0)$ & 52 & $-[\L_2]/3$ & -4 \\
    \hline
    $\Tilde{\sigma_3};~\mathbf{m}=(0,...,3,0)$ & 47 & $(\lz + \lo)/9$ & N/A \\
    \hline
\end{tabular}
\end{center}
\vspace{0.5cm}

\noindent $1^7 \times 4_D$ model with variables $(x_1,...,x_7,y,z)$
\begin{itemize}
    \item $\sigma_1=(12) ~ ; ~ \sigma_2=(12)(34) ~ ; ~ \sigma_3=(12)(34)(56)$
    \item $(T^2)^2 \times (T^2 : \{x_7^3 + y^3 + y z^2 = 0\} \subset \CP^2)/\Z_3 \times \Z_3$
    \item $[pt.]_{T^6}=-\frac{[\Lambda_2]}{3}=-\frac{[\Lambda_0]+[\Lambda_1]}{3}$ (under the assumption of $1_A \times 1_A \cong 4_D$)
\end{itemize}
\begin{center}
\begin{tabular}{|c|c|c|c|}
    \hline
    parity & $h^{2,1}$ modes & O-plane class \\
    \hline
    $\sigma_0;~\mathbf{m}=(0,...,0,+)$ & 84 & $-[\L_0] + [\L_1]$ \\
    \hline
    $\sigma_1;~\mathbf{m}=(0,...,0,+)$ & 63 & $-[\L_2]$ \\
    \hline
    $\sigma_2;~\mathbf{m}=(0,...,0,+)$ & 52 & $([\L_0] - [\L_1])/3$ \\
    \hline
    $\sigma_3;~\mathbf{m}=(0,...,0,+)$ & 47 & $[\L_2]/3$ \\
    \hline
    $\sigma_3;~\mathbf{m}=(0,...,0,-)$ & 44 & $(-[\L_0] + [\L_1])/9$ \\
    \hline
    $\Tilde{\sigma_0};~\mathbf{m}=(0,...,0,+)$ & 84 & $-[\L_0] + [\L_2]$ \\
    \hline
    $\Tilde{\sigma_1};~\mathbf{m}=(0,...,0,+)$ & 63 & $[\L_1]$ \\
    \hline
    $\Tilde{\sigma_2};~\mathbf{m}=(0,...,0,+)$ & 52 & $([\L_0] - [\L_2])/3$ \\
    \hline
    $\Tilde{\sigma_3};~\mathbf{m}=(0,...,0,+)$ & 47 & $-[\L_1]/3$ \\
    \hline
    $\Tilde{\sigma_3};~\mathbf{m}=(0,...,0,-)$ & 44 & $-([\L_0] - [\L_2])/9$ \\
    \hline
\end{tabular}
\end{center}
\vspace{0.5cm}

\noindent $1^5\times4^2_D$ model with variables $(x_1, ..., x_5, y_1, z_1, y_2, z_2)$
\begin{itemize}
    \item $\sigma_1 = (12)_x ~ ; ~ \sigma_2 = (12)_x (34)_x ~ ; ~ \sigma_3 = (12)_x (12)_y (12)_z ~ ; ~ \sigma_4 = (12)_x (34)_x (12)_y (12)_z$
    \item $T^2 \times (T^2 : \{x_i^3 + y_j^3 + y_j z_j^2 = 0\} \subset \CP^2)^2/\Z_3 \times \Z_3$ \\ or $T^2 \times (K3 : \{x_4^3 + x_5^3 + y_1^3 + y_2^3 = 0 ~ ; ~ y_1 z_1^2 + y_2 z_2^2 = 0\} \subset \CP^3 * \CP^1)/\Z_3$
\end{itemize}
\begin{center}
\begin{tabular}{|c|c|c|c|}
    \hline
    parity & $h^{2,1}$ modes & O-plane class \\
    \hline
    $\sigma_0;~\mathbf{m}=(0,...,0,+,+)$ & 84 & $3[\L_1]$ \\
    \hline
    $\sigma_1;~\mathbf{m}=(0,...,0,+,+)$ & 63 & $[\L_0] - [\L_2]$ \\
    \hline
    $\sigma_2;~\mathbf{m}=(0,...,0,+,+)$ & 52 & $-[\L_1]$ \\
    \hline
    $\sigma_3;~\mathbf{m}=(0,...,0,+,+)$ & 47 & $([\L_2] - [\L_0])/3$ \\
    \hline
    $\sigma_4;~\mathbf{m}=(0,...,0,+,+)$ & 44 & $[\L_1]/3$ \\
    \hline
    $\Tilde{\sigma_0};~\mathbf{m}=(0,...,0,+,+)$ & 84 & $-3[\L_0]$ \\
    \hline
    $\Tilde{\sigma_1};~\mathbf{m}=(0,...,0,+,+)$ & 63 & $[\L_1] - [\L_2]$ \\
    \hline
    $\Tilde{\sigma_2};~\mathbf{m}=(0,...,0,+,+)$ & 52 & $[\L_0]$ \\
    \hline
    $\Tilde{\sigma_3};~\mathbf{m}=(0,...,0,+,+)$ & 47 & $([\L_2] - [\L_1])/3$ \\
    \hline
    $\Tilde{\sigma_4};~\mathbf{m}=(0,...,0,+,+)$ & 44 & $-[\L_0]/3$ \\
    \hline
\end{tabular}
\end{center}
\vspace{1.5cm}

\newpage

\noindent $1^5\times4_A\times4_D$ model with variables $(x_1, ..., x_5, y, z, p, q)$
\begin{itemize}
    \item $\sigma_1=(12) ~ ; ~ \sigma_2=(12)(34)$
    \item $T^2 \times (T^2 : \{x_4^3 + y^6 + z^2 = 0\} \subset \WCP^2[2,1,3] \times (T^2 : \{x_5^3 + p^3 + p q^2 = 0\} \subset \CP^2)/\Z_3 \times \Z_3$ \\ or $T^2 \times (K3 : \{x_4^3 + x_5^3 + y^6 + p^3 = 0 ~ ; ~ p q^2 + z^2 = 0\} \subset \WCP^3[2,2,1,2] * \CP^1)/\Z_3$ \footnote{Note that the product ``$*$'' in this equation (and hereafter) is not a direct product. For the equations to define a codimension 2 surface in the ambient space, the rescaling factors in $\mathbb{WCP}^3 [2,2,1,2]$ and $\mathbb{CP}^1$ cannot be independent.}
    \item $[pt.]_{T^6} = \frac{[\L_2]}{3} = \frac{[\L_1]-[\L_0]}{3}$
\end{itemize}
\begin{center}
\begin{tabular}{|c|c|c|c|}
    \hline
    parity & $h^{2,1}$ modes & O-plane class \\
    \hline
    $\sigma_0;~\mathbf{m}=(0,...,0,0,0,+)$ & 63 & $[\L_2] - [\L_0]$ \\
    \hline
    $\sigma_1;~\mathbf{m}=(0,...,0,0,0,+)$ & 52 & $[\L_1]$ \\
    \hline
    $\sigma_2;~\mathbf{m}=(0,...,0,0,0,+)$ & 47 & $([\L_0] - [\L_2])/3$ \\
    \hline
    $\sigma_2;~\mathbf{m}=(0,...,0,0,0,-)$ & 44 & $-[\L_1]/3$ \\
    \hline
    $\sigma_0;~\mathbf{m}=(0,...,0,3,0,+)$ & 63 & $3[\L_1]$ \\
    \hline
    $\sigma_1;~\mathbf{m}=(0,...,0,3,0,+)$ & 52 & $[\L_0]-[\L_2]$ \\
    \hline
    $\sigma_2;~\mathbf{m}=(0,...,0,3,0,+)$ & 47 & $-[\L_1]$ \\
    \hline
    $\sigma_2;~\mathbf{m}=(0,...,0,3,0,-)$ & 44 & $([\L_0]-[\L_2])/3$ \\
    \hline
    $\Tilde{\sigma_0};~\mathbf{m}=(0,...,0,0,0,+)$ & 84 & $-3[\L_0]$ \\
    \hline
    $\Tilde{\sigma_1};~\mathbf{m}=(0,...,0,0,0,+)$ & 63 & $[\L_1] + [\L_2]$ \\
    \hline
    $\Tilde{\sigma_2};~\mathbf{m}=(0,...,0,0,0,+)$ & 52 & $[\L_0]$ \\
    \hline
    $\Tilde{\sigma_2};~\mathbf{m}=(0,...,0,0,0,-)$ & 47 & $-([\L_1] + [\L_2])/3$ \\
    \hline
    $\Tilde{\sigma_0};~\mathbf{m}=(0,...,0,3,0,+)$ & 84 & $[\L_1] + [\L_2]$ \\
    \hline
    $\Tilde{\sigma_1};~\mathbf{m}=(0,...,0,3,0,+)$ & 63 & $[\L_0]$ \\
    \hline
    $\Tilde{\sigma_2};~\mathbf{m}=(0,...,0,3,0,+)$ & 52 & $-([\L_1] + [\L_2])/3$ \\
    \hline
    $\Tilde{\sigma_2};~\mathbf{m}=(0,...,0,3,0,-)$ & 47 & $-[\L_0]/3$ \\
    \hline
\end{tabular}
\end{center}

\newpage

\noindent $1^3\times4_A\times4^2_D$ model with variables $(x_1, x_2, x_3, y, p_1, q_1, p_2, q_2)$
\begin{itemize}
    \item $\sigma_1 = (12)_x ~ ; ~ \sigma_2 = (12)_p (12)_q ~ ; ~ \sigma_3 = (12)_x (12)_p (12)_q$
    \item $(T^2 : \{x_1^3 + y^6 + z^2 = 0\} \subset \WCP^2[2,1,3]) \times (T^2 : \{x_i^3 + p_j^3 + p_j q_j^2 = 0\} \subset \CP^2)/\Z_3 \times \Z_3$ \\ or $T^2 \times (K3 : \{y^6 + p_1^3 + p_2^3 + z^2 = 0 ~ ; ~ p_1 q_1^2 + p_2 q_2^2 = 0\} \subset \WCP^3[1,2,2,3] \times \CP^1)/\Z_3$
\end{itemize}
\begin{center}
\begin{tabular}{|c|c|c|c|}
    \hline
    parity & $h^{2,1}$ modes & O-plane class  \\
    \hline
    $\sigma_0;~\mathbf{m}=(0,...,0,+,+)$ & 63 & $3[\L_2]$ \\
    \hline
    $\sigma_0;~\mathbf{m}=(0,...,3,+,+)$ & 63 & $-3([\L_0] + [\L_1])$ \\
    \hline
    $\sigma_1;~\mathbf{m}=(0,...,0,+,+)$ & 52 & $[\L_0] + [\L_1]$ \\
    \hline
    $\sigma_1;~\mathbf{m}=(0,...,3,+,+)$ & 52 & $-3[\L_2]$ \\
    \hline
    $\sigma_2;~\mathbf{m}=(0,...,0,+,+)$ & 47 & $-[\L_2]$ \\
    \hline
    $\sigma_2;~\mathbf{m}=(0,...,3,+,+)$ & 47 & $-[\L_0] - [\L_1]$ \\
    \hline
    $\sigma_3;~\mathbf{m}=(0,...,0,+,+)$ & 44 & $-([\L_0] + [\L_1])/3$ \\
    \hline
    $\sigma_3;~\mathbf{m}=(0,...,3,+,+)$ & 44 & $[\L_2]$ \\
    \hline
    $\Tilde{\sigma_0};~\mathbf{m}=(0,...,0,+,+)$ & 84 & $3([\L_2] - [\L_0])$ \\
    \hline
    $\Tilde{\sigma_0};~\mathbf{m}=(0,...,3,+,+)$ & 84 & $-3[\L_1]$ \\
    \hline
    $\Tilde{\sigma_1};~\mathbf{m}=(0,...,0,+,+)$ & 63 & $3[\L_1]$ \\
    \hline
    $\Tilde{\sigma_1};~\mathbf{m}=(0,...,3,+,+)$ & 63 & $-[\L_0] + [\L_2] $ \\
    \hline
    $\Tilde{\sigma_2};~\mathbf{m}=(0,...,0,+,+)$ & 52 & $[\L_0] - [\L_2]$ \\
    \hline
    $\Tilde{\sigma_2};~\mathbf{m}=(0,...,3,+,+)$ & 52 & $[\L_1]$ \\
    \hline
    $\Tilde{\sigma_3};~\mathbf{m}=(0,...,0,+,+)$ & 47 & $-[\L_1]$ \\
    \hline
    $\Tilde{\sigma_3};~\mathbf{m}=(0,...,3,+,+)$ & 47 & $([\L_2] - [\L_0])/3$ \\
    \hline
\end{tabular}
\end{center}

\noindent $1\times4^4_D$ model with variables $(x, y_1, z_1, y_2, z_2, y_3, z_3, y_4, z_4)$
\begin{itemize}
    \item $\sigma_1 = (12)_y (12)_z ~ ; ~ \sigma_2 = (12)_y (12)_z (34)_y (34)_z$
    \item $(T^2 : \{x^3 + y_1^3 + y_1 z_1^2 = 0\} \subset \CP^2) \times (K3 : \{y_2^3 + y_3^3 + y_4^3 + y_2 z_2^2 = 0 ~ ; ~ y_3 z_3^2 + y_4 z_4^2 = 0\} \subset \CP^3 \times \CP^1)/\Z_3$
\end{itemize}
\begin{center}
\begin{tabular}{|c|c|c|c|}
    \hline
    parity & $h^{2,1}$ modes & O-plane class \\
    \hline
    $\sigma_0;~\mathbf{m}=(0,0,0,0,0,+,+,+,+)$ & 84 & $-9\lt$ \\
    \hline
    $\sigma_0;~\mathbf{m}=(0,0,0,0,0,+,+,+,-)$ & 63 & $3(\lz - \lo)$ \\
    \hline
    $\sigma_1;~\mathbf{m}=(0,0,0,0,0,+,+,+,+)$ & 52 & $3\lt$ \\
    \hline
    $\sigma_1;~\mathbf{m}=(0,0,0,0,0,+,+,+,-)$ & 47 & $\lo - \lz$ \\
    \hline
    $\sigma_2;~\mathbf{m}=(0,0,0,0,0,+,+,+,+)$ & 44 & $-\lt$ \\
    \hline
    $\Tilde{\sigma_0};~\mathbf{m}=(0,0,0,0,0,+,+,+,+)$ & 84 & $9\lo$ \\
    \hline
    $\Tilde{\sigma_0};~\mathbf{m}=(0,0,0,0,0,+,+,+,-)$ & 63 & $3(\lz - \lt)$ \\
    \hline
    $\Tilde{\sigma_1};~\mathbf{m}=(0,0,0,0,0,+,+,+,+)$ & 52 & $-3\lo$ \\
    \hline
    $\Tilde{\sigma_1};~\mathbf{m}=(0,0,0,0,0,+,+,+,-)$ & 47 & $\lt - \lz$ \\
    \hline
    $\Tilde{\sigma_2};~\mathbf{m}=(0,0,0,0,0,+,+,+,+)$ & 44 & $\lo$ \\
    \hline
\end{tabular}
\end{center}

\noindent $1\times4^3_A\times4_D$ model with variables $(x, y_1, y_2, y_3, p, q)$
\begin{itemize}
    \item $\sigma_1 = (12)_y$
    \item $(T^2 : \{x^3 + y_1^6 + z^2 = 0 \} \subset \WCP^2[2,1,3]) \times (K3 : \{y_2^6 + y_3^6 + p^3 + p q^2 = 0\} \subset \WCP^3[1,1,2,2])/\Z_3$ \\ or $(T^2 : \{x^3 + p^3 + p q^2 = 0\} \subset \CP^2) \times (K3 : \{y_1^6 + y_2^6 + y_3^6 + z^2 = 0\} \subset \WCP^3[1,1,1,3])/\Z_3$
\end{itemize}
\begin{center}
\begin{tabular}{|c|c|c|c|}
    \hline
    parity & $h^{2,1}$ modes & O-plane class \\
    \hline
    $\sigma_0;~\mathbf{m}=(0,0,0,0,0,+)$ & 83 & $9\lo$ \\
    \hline
    $\sigma_0;~\mathbf{m}=(0,0,3,3,3,+)$ & 83 & $5(-\lz + \lt)$ \\
    \hline
    $\sigma_0;~\mathbf{m}=(0,0,0,0,0,-)$ & 62 & $3(\lz - \lt)$ \\
    \hline
    $\sigma_0;~\mathbf{m}=(0,0,3,3,3,-)$ & 62 & $-5\lo$ \\
    \hline
    $\sigma_1;~\mathbf{m}=(0,0,0,0,0,+)$ & 51 & $\lz - \lt$ \\
    \hline
    $\sigma_1;~\mathbf{m}=(0,0,0,3,0,+)$ & 51 & $-3\lo$ \\
    \hline
    $\sigma_1;~\mathbf{m}=(0,0,0,0,0,-)$ & 46 & $-\lo$ \\
    \hline
    $\sigma_1;~\mathbf{m}=(0,0,0,3,0,-)$ & 46 & $\lt - \lz$ \\
    \hline
    $\Tilde{\sigma_0};~\mathbf{m}=(0,0,0,0,0,+)$ & 84 & $5(\lo + \lt)$ \\
    \hline
    $\Tilde{\sigma_0};~\mathbf{m}=(0,0,3,3,3,+)$ & 84 & $9\lz$ \\
    \hline
    $\Tilde{\sigma_0};~\mathbf{m}=(0,0,0,0,0,-)$ & 63 & $5\lz$ \\
    \hline
    $\Tilde{\sigma_0};~\mathbf{m}=(0,0,3,3,3,-)$ & 63 & $3(\lo + \lt)$ \\
    \hline
    $\Tilde{\sigma_1};~\mathbf{m}=(0,0,0,0,0,+)$ & 52 & $3\lz$ \\
    \hline
    $\Tilde{\sigma_1};~\mathbf{m}=(0,0,0,3,0,+)$ & 52 & $\lo + \lt$ \\
    \hline
    $\Tilde{\sigma_1};~\mathbf{m}=(0,0,0,0,0,-)$ & 47 & $-(\lo + \lt)$ \\
    \hline
    $\Tilde{\sigma_1};~\mathbf{m}=(0,0,0,3,0,-)$ & 47 & $\lz$ \\
    \hline
\end{tabular}
\end{center}

\newpage

\noindent $1\times4_A\times4^3_D$ model with variables $(x, y, p_1, q_1, p_2, q_2, p_3, q_3, z)$
\begin{itemize}
    \item $\sigma_1 = (12)_p (12)_q$
    \item $(T^2 : \{x^3 + y^6 + z^2 = 0 \} \subset \WCP^2[2,1,3])$ \\ \hspace*{\fill} $\times  (K3 : \{p_1^3 + p_2^3 + p_3^3 + p_1 q_1^2 = 0 ~ ; ~ p_2 q_2^2 + p_3 q_3^2 = 0\} \subset \CP^3 \times \CP^1)/\Z_3$ \\ or $(T^2 : \{x^3 + p_1^3 + p_1 q_1^2 = 0\} \subset \CP^2)$ \\ \hspace*{\fill} $\times (K3 : \{y^6 + p_2^3 + p_3^3 + z^2 = 0 ~ ; ~ p_2 q_2^2 + p_3 q_3^2 = 0\} \subset \WCP^3[1,2,2,3] \times \CP^1)/\Z_3$
\end{itemize}
\begin{center}
\begin{tabular}{|c|c|c|c|}
    \hline
    parity & $h^{2,1}$ modes & O-plane class \\
    \hline
    $\sigma_0;~\mathbf{m}=(0,...,0,+,+,+)$ & 63 & $3(\lo + \lt)$ \\
    \hline
    $\sigma_0;~\mathbf{m}=(0,...,3,+,+,+)$ & 63 & $9\lz$ \\
    \hline
    $\sigma_0;~\mathbf{m}=(0,...,0,+,+,-)$ & 52 & $3\lz$ \\
    \hline
    $\sigma_0;~\mathbf{m}=(0,...,3,+,+,-)$ & 52 & $-3(\lo + \lt)$ \\
    \hline
    $\sigma_1;~\mathbf{m}=(0,...,0,+,+,+)$ & 47 & $-(\lo + \lt)$ \\
    \hline
    $\sigma_1;~\mathbf{m}=(0,...,3,+,+,+)$ & 47 & $-3\lz$ \\
    \hline
    $\sigma_1;~\mathbf{m}=(0,...,0,+,+,-)$ & 44 & $-\lz$ \\
    \hline
    $\sigma_1;~\mathbf{m}=(0,...,3,+,+,-)$ & 44 & $(\lo + \lt)$ \\
    \hline
    $\Tilde{\sigma_0};~\mathbf{m}=(0,...,0,+,+,+)$ & 84 & $9\lt$ \\
    \hline
    $\Tilde{\sigma_0};~\mathbf{m}=(0,...,3,+,+,+)$ & 84 & $-3(\lz + \lo)$ \\
    \hline
    $\Tilde{\sigma_0};~\mathbf{m}=(0,...,0,+,+,-)$ & 63 & $3(\lz + \lo)$ \\
    \hline
    $\Tilde{\sigma_0};~\mathbf{m}=(0,...,3,+,+,-)$ & 63 & $3\lt$ \\
    \hline
    $\Tilde{\sigma_1};~\mathbf{m}=(0,...,0,+,+,+)$ & 52 & $-3\lt$ \\
    \hline
    $\Tilde{\sigma_1};~\mathbf{m}=(0,...,3,+,+,+)$ & 52 & $-\lz + \lo$ \\
    \hline
    $\Tilde{\sigma_1};~\mathbf{m}=(0,...,0,+,+,-)$ & 47 & $\lz - \lo$ \\
    \hline
    $\Tilde{\sigma_1};~\mathbf{m}=(0,...,3,+,+,-)$ & 47 & $\lt$ \\
    \hline
\end{tabular}
\end{center}
\vspace{0.5cm}

\printbibliography

@article{Gepner:1987qi,
    author = "Gepner, Doron",
    editor = "Schellekens, B.",
    title = "{Space-Time Supersymmetry in Compactified String Theory and Superconformal Models}",
    reportNumber = "Print-87-0370 (PRINCETON), PUPT-1056",
    doi = "10.1016/0550-3213(88)90397-5",
    journal = "Nucl. Phys. B",
    volume = "296",
    pages = "757",
    year = "1988"
}

@article{Gepner:1987vz,
    author = "Gepner, Doron",
    title = "{Exactly Solvable String Compactifications on Manifolds of SU(N) Holonomy}",
    reportNumber = "PRINT-87-0727 (PRINCETON), PUPT-1066",
    doi = "10.1016/0370-2693(87)90938-5",
    journal = "Phys. Lett. B",
    volume = "199",
    pages = "380--388",
    year = "1987"
}

@article{Brunner_2007,
   title={Orientifolds of Gepner models},
   volume={2007},
   ISSN={1029-8479},
   url={http://dx.doi.org/10.1088/1126-6708/2007/02/001},
   DOI={10.1088/1126-6708/2007/02/001},
   number={02},
   journal={Journal of High Energy Physics},
   publisher={Springer Science and Business Media LLC},
   author={Brunner, Ilka and Hori, Kentaro and Hosomichi, Kazuo and Walcher, Johannes},
   year={2007},
   month=feb, pages={001–001} }

@book{Blumenhagen:2013fgp,
    author = {Blumenhagen, Ralph and L{\"u}st, Dieter and Theisen, Stefan},
    title = "{Basic concepts of string theory}",
    doi = "10.1007/978-3-642-29497-6",
    isbn = "978-3-642-29496-9",
    publisher = "Springer",
    address = "Heidelberg, Germany",
    series = "Theoretical and Mathematical Physics",
    year = "2013"
}

@article{CAPPELLI1987445,
title = {Modular invariant partition functions in two dimensions},
journal = {Nuclear Physics B},
volume = {280},
pages = {445-465},
year = {1987},
issn = {0550-3213},
author = {A. Cappelli and C. Itzykson and J.-B. Zuber},
}

@article{Fuchs:1989yv,
    author = "Fuchs, Jurgen and Klemm, Albrecht and Scheich, Christoph and Schmidt, Michael G.",
    title = "{Spectra and Symmetries of Gepner Models Compared to Calabi-yau Compactifications}",
    reportNumber = "HD-THEP-89-25",
    doi = "10.1016/0003-4916(90)90119-9",
    journal = "Annals Phys.",
    volume = "204",
    pages = "1--51",
    year = "1990"
}

@article{Qiu:1987ux,
    author = "Qiu, Zong-an",
    title = "{Modular Invariant Partition Functions for $N=2$ Superconformal Field Theories}",
    reportNumber = "IASSNS-HEP-87-26",
    doi = "10.1016/0370-2693(87)90906-3",
    journal = "Phys. Lett. B",
    volume = "198",
    pages = "497--502",
    year = "1987"
}

@article{Vafa:1988uu,
    author = "Vafa, Cumrun and Warner, Nicholas P.",
    title = "{Catastrophes and the Classification of Conformal Theories}",
    reportNumber = "HUTP-88-A037",
    doi = "10.1016/0370-2693(89)90473-5",
    journal = "Phys. Lett. B",
    volume = "218",
    pages = "51--58",
    year = "1989"
}

@article{Greene:1988ut,
    author = "Greene, Brian R. and Vafa, C. and Warner, N. P.",
    title = "{Calabi-Yau Manifolds and Renormalization Group Flows}",
    reportNumber = "HUTP-88-A047A",
    doi = "10.1016/0550-3213(89)90471-9",
    journal = "Nucl. Phys. B",
    volume = "324",
    pages = "371",
    year = "1989"
}

@article{Brunner_2004,
   title={Orientifolds and Mirror Symmetry},
   volume={2004},
   ISSN={1029-8479},
   url={http://dx.doi.org/10.1088/1126-6708/2004/11/005},
   DOI={10.1088/1126-6708/2004/11/005},
   number={11},
   journal={Journal of High Energy Physics},
   publisher={Springer Science and Business Media LLC},
   author={Brunner, Ilka and Hori, Kentaro},
   year={2004},
   month=nov, pages={005–005} }

@article{Becker_2007,
    author = "Becker, Katrin and Becker, Melanie and Vafa, Cumrun and Walcher, Johannes",
    title = "{Moduli Stabilization in Non-Geometric Backgrounds}",
    eprint = "hep-th/0611001",
    archivePrefix = "arXiv",
    reportNumber = "HUTP-06-A044",
    doi = "10.1016/j.nuclphysb.2007.01.034",
    journal = "Nucl. Phys. B",
    volume = "770",
    pages = "1--46",
    year = "2007"
}

@article{Brunner_2004_RCFT,
   title={Notes on Orientifolds of Rational Conformal Field Theories},
   volume={2004},
   ISSN={1029-8479},
   url={http://dx.doi.org/10.1088/1126-6708/2004/07/023},
   DOI={10.1088/1126-6708/2004/07/023},
   number={07},
   journal={Journal of High Energy Physics},
   publisher={Springer Science and Business Media LLC},
   author={Brunner, Ilka and Hori, Kentaro},
   year={2004},
   month=jul, pages={023–023} }

@article{Walcher_2005,
    author = "Walcher, Johannes",
    title = "{Stability of Landau-Ginzburg branes}",
    eprint = "hep-th/0412274",
    archivePrefix = "arXiv",
    doi = "10.1063/1.2007590",
    journal = "J. Math. Phys.",
    volume = "46",
    pages = "082305",
    year = "2005"
}

@article{Hori_2008,
    author = "Hori, Kentaro and Walcher, Johannes",
    title = "{D-brane Categories for Orientifolds: The Landau-Ginzburg Case}",
    eprint = "hep-th/0606179",
    archivePrefix = "arXiv",
    doi = "10.1088/1126-6708/2008/04/030",
    journal = "JHEP",
    volume = "04",
    pages = "030",
    year = "2008"
}

@article{Kachru_2003,
    author = "Kachru, Shamit and Schulz, Michael B. and Trivedi, Sandip",
    title = "{Moduli stabilization from fluxes in a simple IIB orientifold}",
    eprint = "hep-th/0201028",
    archivePrefix = "arXiv",
    reportNumber = "SLAC-PUB-9066, SU-ITP-01-49, TIFR-TH-01-51",
    doi = "10.1088/1126-6708/2003/10/007",
    journal = "JHEP",
    volume = "10",
    pages = "007",
    year = "2003"
}

@article{Becker:2024ayh,
    author = "Becker, Katrin and Brady, Nathan and Gra{\~n}a, Mariana and Morros, Miguel and Sengupta, Anindya and You, Qi",
    title = "{Tadpole conjecture in non-geometric backgrounds}",
    eprint = "2407.16758",
    archivePrefix = "arXiv",
    primaryClass = "hep-th",
    doi = "10.1007/JHEP10(2024)021",
    journal = "JHEP",
    volume = "10",
    pages = "021",
    year = "2024"
}

@article{Frey_2002,
    author = "Frey, Andrew R. and Polchinski, Joseph",
    title = "{N=3 warped compactifications}",
    eprint = "hep-th/0201029",
    archivePrefix = "arXiv",
    reportNumber = "NSF-ITP-01-77",
    doi = "10.1103/PhysRevD.65.126009",
    journal = "Phys. Rev. D",
    volume = "65",
    pages = "126009",
    year = "2002"
}

@article{Lynker:1990gh,
    author = "Lynker, Monika and Schimmrigk, Rolf",
    title = "{A-D-E quantum Calabi-Yau manifolds}",
    reportNumber = "NSF-ITP-89-182, UTTG-42-89",
    doi = "10.1016/0550-3213(90)90536-M",
    journal = "Nucl. Phys. B",
    volume = "339",
    pages = "121--157",
    year = "1990"
}

@article{Blumenhagen_2003,
   title={Moduli stabilization in chiral type IIB orientifold models with fluxes},
   volume={663},
   ISSN={0550-3213},
   url={http://dx.doi.org/10.1016/S0550-3213(03)00392-4},
   DOI={10.1016/s0550-3213(03)00392-4},
   number={1–2},
   journal={Nuclear Physics B},
   publisher={Elsevier BV},
   author={Blumenhagen, Ralph and Lüst, Dieter and Taylor, Tomasz R.},
   year={2003},
   month=jul, pages={319–342} }

@article{Blumenhagen_2005,
   title={Chiral D-brane Models with Frozen Open String Moduli},
   volume={2005},
   ISSN={1029-8479},
   url={http://dx.doi.org/10.1088/1126-6708/2005/03/050},
   DOI={10.1088/1126-6708/2005/03/050},
   number={03},
   journal={Journal of High Energy Physics},
   publisher={Springer Science and Business Media LLC},
   author={Blumenhagen, Ralph and Cvetic, Mirjam and Marchesano, Fernando and Shiu, Gary},
   year={2005},
   month=mar, pages={050–050} }

@misc{antoniadis2024fluxvacuatypeiib,
    author = "Antoniadis, Ignatios and Guillen, Anthony and Lacombe, Osmin",
    title = "{Flux vacua in type IIB compactifications on orbifolds: their finiteness and minimal string coupling}",
    eprint = "2404.18995",
    archivePrefix = "arXiv",
    primaryClass = "hep-th",
    doi = "10.1007/JHEP09(2024)016",
    journal = "JHEP",
    volume = "09",
    pages = "016",
    year = "2024"
}

@misc{becker2023fluxes19landauginzburgmodel,
      title={On Fluxes in the $1^9$ Landau-Ginzburg Model}, 
      author={Katrin Becker and Nathan Brady and Anindya Sengupta},
      year={2023},
      eprint={2310.00770},
      archivePrefix={arXiv},
      primaryClass={hep-th},
      url={https://arxiv.org/abs/2310.00770}, 
}

@article{Kachru_2002,
    author = "Kachru, Shamit and Pearson, John and Verlinde, Herman L.",
    title = "{Brane / flux annihilation and the string dual of a nonsupersymmetric field theory}",
    eprint = "hep-th/0112197",
    archivePrefix = "arXiv",
    reportNumber = "SLAC-PUB-9083, SU-ITP-01-50, PUPT-2019",
    doi = "10.1088/1126-6708/2002/06/021",
    journal = "JHEP",
    volume = "06",
    pages = "021",
    year = "2002"
}

@article{Becker_2024_higherorder,
   title={Stabilizing massless fields with fluxes in Landau-Ginzburg models},
   volume={2024},
   ISSN={1029-8479},
   url={http://dx.doi.org/10.1007/JHEP08(2024)069},
   DOI={10.1007/jhep08(2024)069},
   number={8},
   journal={Journal of High Energy Physics},
   publisher={Springer Science and Business Media LLC},
   author={Becker, Katrin and Rajaguru, Muthusamy and Sengupta, Anindya and Walcher, Johannes and Wrase, Timm},
   year={2024},
   month=aug }

@misc{chen2025symmetriesmtheorylikevacuadimensions,
      title={Symmetries and M-theory-like Vacua in Four Dimensions}, 
      author={Shi Chen and Damian van de Heisteeg and Cumrun Vafa},
      year={2025},
      eprint={2503.16599},
      archivePrefix={arXiv},
      primaryClass={hep-th},
      url={https://arxiv.org/abs/2503.16599}, 
}

@article{Kapustin_2004,
   author = "Kapustin, Anton and Li, Yi",
    title = "{D-branes in topological minimal models: The Landau-Ginzburg approach}",
    eprint = "hep-th/0306001",
    archivePrefix = "arXiv",
    doi = "10.1088/1126-6708/2004/07/045",
    journal = "JHEP",
    volume = "07",
    pages = "045",
    year = "2004"
}

@article{Kapustin_2003,
   author = "Kapustin, Anton and Li, Yi",
    title = "{D branes in Landau-Ginzburg models and algebraic geometry}",
    eprint = "hep-th/0210296",
    archivePrefix = "arXiv",
    reportNumber = "CALT-68-2412",
    doi = "10.1088/1126-6708/2003/12/005",
    journal = "JHEP",
    volume = "12",
    pages = "005",
    year = "2003"
}

@misc{kapustin2003topologicalcorrelatorslandauginzburgmodels,
      title={Topological Correlators in Landau-Ginzburg Models with Boundaries}, 
      author={Anton Kapustin and Yi Li},
      year={2003},
      eprint={hep-th/0305136},
      archivePrefix={arXiv},
      primaryClass={hep-th},
      url={https://arxiv.org/abs/hep-th/0305136}, 
}

@article{Kapustin:2003ga,
    author = "Kapustin, Anton and Li, Yi",
    title = "{Topological correlators in Landau-Ginzburg models with boundaries}",
    eprint = "hep-th/0305136",
    archivePrefix = "arXiv",
    doi = "10.4310/ATMP.2003.v7.n4.a5",
    journal = "Adv. Theor. Math. Phys.",
    volume = "7",
    number = "4",
    pages = "727--749",
    year = "2003"
}

@article{Vafa:1986wx,
    author = "Vafa, Cumrun",
    title = "{Modular Invariance and Discrete Torsion on Orbifolds}",
    reportNumber = "HUTP-86/A011",
    doi = "10.1016/0550-3213(86)90379-2",
    journal = "Nucl. Phys. B",
    volume = "273",
    pages = "592--606",
    year = "1986"
}

@article{Brunner_2005,
    author = "Brunner, Ilka and Gaberdiel, Matthias R.",
    title = "{Matrix factorisations and permutation branes}",
    eprint = "hep-th/0503207",
    archivePrefix = "arXiv",
    doi = "10.1088/1126-6708/2005/07/012",
    journal = "JHEP",
    volume = "07",
    pages = "012",
    year = "2005"
}

@article{Douglas_2001,
    author = "Douglas, Michael R.",
    title = "{D-branes, categories and N=1 supersymmetry}",
    eprint = "hep-th/0011017",
    archivePrefix = "arXiv",
    reportNumber = "RUNHETC-2000-42",
    doi = "10.1063/1.1374448",
    journal = "J. Math. Phys.",
    volume = "42",
    pages = "2818--2843",
    year = "2001"
}

@inproceedings{ASPINWALL_2005,
   title={D-Branes on Calabi-Yau Manifolds},
   url={http://dx.doi.org/10.1142/9789812775108_0001},
   DOI={10.1142/9789812775108_0001},
   booktitle={Progress In String Theory},
   publisher={World Scientific},
   author={Aspinwall, PAUL S.},
   year={2005},
   month=jul, pages={1–152} }

@article{Hori_2004,
    author = "Hori, Kentaro and Walcher, Johannes",
    title = "{D-branes from matrix factorizations}",
    eprint = "hep-th/0409204",
    archivePrefix = "arXiv",
    doi = "10.1016/j.crhy.2004.09.016",
    journal = "Comptes Rendus Physique",
    volume = "5",
    pages = "1061--1070",
    year = "2004"
}

@article{Gukov:1999ya,
    author = "Gukov, Sergei and Vafa, Cumrun and Witten, Edward",
    title = "{CFT's from Calabi-Yau four folds}",
    eprint = "hep-th/9906070",
    archivePrefix = "arXiv",
    reportNumber = "HUTP-99-A034, IASSNS-HEP-99-52, PUPT-1864",
    doi = "10.1016/S0550-3213(00)00373-4",
    journal = "Nucl. Phys. B",
    volume = "584",
    pages = "69--108",
    year = "2000",
    note = "[Erratum: Nucl.Phys.B 608, 477--478 (2001)]"
}

@article{Plauschinn:2018wbo,
    author = "Plauschinn, Erik",
    title = "{Non-geometric backgrounds in string theory}",
    eprint = "1811.11203",
    archivePrefix = "arXiv",
    primaryClass = "hep-th",
    reportNumber = "LMU-ASC 79/18",
    doi = "10.1016/j.physrep.2018.12.002",
    journal = "Phys. Rept.",
    volume = "798",
    pages = "1--122",
    year = "2019"
}

@article{Balasubramanian:2005zx,
    author = "Balasubramanian, Vijay and Berglund, Per and Conlon, Joseph P. and Quevedo, Fernando",
    title = "{Systematics of moduli stabilisation in Calabi-Yau flux compactifications}",
    eprint = "hep-th/0502058",
    archivePrefix = "arXiv",
    reportNumber = "DAMTP-2005-10, UNH-05-01, UPR-1109-T",
    doi = "10.1088/1126-6708/2005/03/007",
    journal = "JHEP",
    volume = "03",
    pages = "007",
    year = "2005"
}

@article{Blumenhagen:2016axv,
    author = "Blumenhagen, Ralph and Fuchs, Michael and Plauschinn, Erik",
    title = "{Partial SUSY Breaking for Asymmetric Gepner Models and Non-geometric Flux Vacua}",
    eprint = "1608.00595",
    archivePrefix = "arXiv",
    primaryClass = "hep-th",
    doi = "10.1007/JHEP01(2017)105",
    journal = "JHEP",
    volume = "01",
    pages = "105",
    year = "2017"
}

@article{Blumenhagen:2016rof,
    author = "Blumenhagen, Ralph and Fuchs, Michael and Plauschinn, Erik",
    title = "{The Asymmetric CFT Landscape in D=4,6,8 with Extended Supersymmetry}",
    eprint = "1611.04617",
    archivePrefix = "arXiv",
    primaryClass = "hep-th",
    doi = "10.1002/prop.201700006",
    journal = "Fortsch. Phys.",
    volume = "65",
    number = "3-4",
    pages = "1700006",
    year = "2017"
}

@article{Witten:1993yc,
    author = "Witten, Edward",
    editor = "Greene, B. and Yau, Shing-Tung",
    title = "{Phases of N=2 theories in two-dimensions}",
    eprint = "hep-th/9301042",
    archivePrefix = "arXiv",
    reportNumber = "IASSNS-HEP-93-3",
    doi = "10.1016/0550-3213(93)90033-L",
    journal = "Nucl. Phys. B",
    volume = "403",
    pages = "159--222",
    year = "1993"
}

@article{Gu:2023hgm,
    author = "Gu, Wei and Melnikov, Ilarion V. and Sharpe, Eric",
    title = "{Quantum cohomology from mixed Higgs-Coulomb phases}",
    eprint = "2308.12334",
    archivePrefix = "arXiv",
    primaryClass = "hep-th",
    doi = "10.1007/JHEP02(2024)010",
    journal = "JHEP",
    volume = "02",
    pages = "010",
    year = "2024"
}

@article{Sethi:1994ch,
    author = "Sethi, S.",
    editor = "Greene, B. and Yau, Shing-Tung",
    title = "{Supermanifolds, rigid manifolds and mirror symmetry}",
    eprint = "hep-th/9404186",
    archivePrefix = "arXiv",
    reportNumber = "HUTP-94-A002",
    doi = "10.1016/0550-3213(94)90649-1",
    journal = "Nucl. Phys. B",
    volume = "430",
    pages = "31--50",
    year = "1994"
}

@article{FUCHS1989317,
title = {Gepner models with arbitrary affine invariants and the associated Calabi-Yau spaces},
journal = {Physics Letters B},
volume = {232},
number = {3},
pages = {317-322},
year = {1989},
issn = {0370-2693},
doi = {https://doi.org/10.1016/0370-2693(89)90750-8},
url = {https://www.sciencedirect.com/science/article/pii/0370269389907508},
author = {Jürgen Fuchs and Albrecht Klemm and Christoph Scheich and Michael G. Schmidt}
}

@article{Braun:2005eg,
    author = "Braun, Volker and Schafer-Nameki, Sakura",
    title = "{D-brane charges in Gepner models}",
    eprint = "hep-th/0511100",
    archivePrefix = "arXiv",
    reportNumber = "DESY-05-227, UPR-1137-T, ZMP-HH-05-20",
    doi = "10.1063/1.2245211",
    journal = "J. Math. Phys.",
    volume = "47",
    pages = "092304",
    year = "2006"
}

@article{Witten:1998cd,
    author = "Witten, Edward",
    title = "{D-branes and K-theory}",
    eprint = "hep-th/9810188",
    archivePrefix = "arXiv",
    reportNumber = "IASSNS-HEP-98-82",
    doi = "10.1088/1126-6708/1998/12/019",
    journal = "JHEP",
    volume = "12",
    pages = "019",
    year = "1998"
}

@article{Bena:2020xrh,
    author = "Bena, Iosif and Bl{\r{a}}b{\"a}ck, Johan and Gra{\~n}a, Mariana and L{\"u}st, Severin",
    title = "{The tadpole problem}",
    eprint = "2010.10519",
    archivePrefix = "arXiv",
    primaryClass = "hep-th",
    doi = "10.1007/JHEP11(2021)223",
    journal = "JHEP",
    volume = "11",
    pages = "223",
    year = "2021"
}

@article{Ishiguro:2024coq,
    author = "Ishiguro, Keiya and Kai, Takafumi and Otsuka, Hajime",
    title = "{Stabilization of a twisted modulus on a mirror of rigid Calabi-Yau manifold}",
    eprint = "2406.08970",
    archivePrefix = "arXiv",
    primaryClass = "hep-th",
    reportNumber = "KEK-TH-2627, KYUSHU-HET-291",
    doi = "10.1007/JHEP10(2024)060",
    journal = "JHEP",
    volume = "10",
    pages = "060",
    year = "2024"
}

@article{Grimm:2024fip,
    author = "Grimm, Thomas W. and van de Heisteeg, Damian",
    title = "{Exact flux vacua, symmetries, and the structure of the landscape}",
    eprint = "2404.12422",
    archivePrefix = "arXiv",
    primaryClass = "hep-th",
    doi = "10.1007/JHEP01(2025)005",
    journal = "JHEP",
    volume = "01",
    pages = "005",
    year = "2025"
}

@article{Mohseni:2025tig,
    author = "Mohseni, Amineh and Vafa, Cumrun",
    title = "{Symmetry points of $ \mathcal{N}=1 $ modular geometry}",
    eprint = "2510.19927",
    archivePrefix = "arXiv",
    primaryClass = "hep-th",
    doi = "10.1007/JHEP02(2026)202",
    journal = "JHEP",
    volume = "02",
    pages = "202",
    year = "2026"
}

@article{Lust:2022mhk,
    author = {L{\"u}st, Severin and Wiesner, Max},
    title = "{The tadpole conjecture in the interior of moduli space}",
    eprint = "2211.05128",
    archivePrefix = "arXiv",
    primaryClass = "hep-th",
    doi = "10.1007/JHEP12(2023)029",
    journal = "JHEP",
    volume = "12",
    pages = "029",
    year = "2023"
}

@article{Cascales:2003zp,
    author = "Cascales, Juan F. G. and Uranga, Angel M.",
    title = "{Chiral 4d string vacua with D branes and NSNS and RR fluxes}",
    eprint = "hep-th/0303024",
    archivePrefix = "arXiv",
    reportNumber = "FTUAM-03-03, IFT-UAM-CSIC-03-07",
    doi = "10.1088/1126-6708/2003/05/011",
    journal = "JHEP",
    volume = "05",
    pages = "011",
    year = "2003"
}
\end{document}